\documentclass[twocolumn,tighten,times]{aastex63}
\usepackage{apjfonts}

\newcommand{\caat}{$Ca_{\rm AT}$}
\newcommand{\cactio}{$Ca_{\rm CTIO}$}

\newcommand{\cnjwl}{$cn_{\rm JWL}$}
\newcommand{\cnpjwl}{$cn^\prime_{\rm JWL}$}
\newcommand{\cnw}{CN-w}
\newcommand{\cns}{CN-s}

\newcommand{\cnwp}{CN-w$_{\rm MP}$}
\newcommand{\cnwr}{CN-w$_{\rm MR}$}
\newcommand{\cnsp}{CN-s$_{\rm MP}$}
\newcommand{\cnsr}{CN-s$_{\rm MR}$}
\newcommand{\cnwhb}{CN-w(RHB)}

\newcommand{\nrgb}{$n$(\cnw):$n$(\cns)}

\newcommand{\pby}{$\parallel$$(b-y)$}

\newcommand{\pcnjwl}{$\parallel$$cn_{\rm JWL}$}
\newcommand{\pcnpjwl}{$\parallel$$cn^\prime_{\rm JWL}$}

\newcommand{\phkjwl}{$\parallel$$hk_{\rm JWL}$}

\newcommand{\pcy}{$\parallel$$cy$}
\newcommand{\pmone}{$\parallel$$m1$}

\newcommand{\vbump}{$V_{\rm bump}$}

\newcommand{\vvhb}{$V - V_{\rm HB}$}

\newcommand{\cnwave}{$\lambda$3883}

\newcommand{\dy}{$\Delta Y$}

\newcommand{\hkjwl}{$hk_{\rm JWL}$}

\newcommand{\mfeh}{$\langle$[Fe/H]$\rangle$}

\newcommand{\fehhk}{[Fe/H]$_{hk}$}

\newcommand{\ofe}{[O/Fe]}
\newcommand{\nafe}{[Na/Fe]}

\newcommand{\feh}{[Fe/H]}

\newcommand{\dfeh}{$\Delta$[Fe/H]}
\newcommand{\dnfe}{$\Delta$[N/Fe]}
\newcommand{\dcfe}{$\Delta$[C/Fe]}
\newcommand{\msun}{$M_{\rm \odot}$}
\newcommand{\cubi}{$\Delta C_{UBI}$}
\newcommand{\nrgbbv}{$\Delta(B-V)_{\rm blue}:\Delta(B-V)_{\rm red}$}
\newcommand{\vbumpto}{$V_{\rm bump}-V_{\rm TO}$}

\shorttitle{47 Tuc}
\shortauthors{Lee}

\begin{document}

\title{Multiple Stellar Populations of Globular Clusters From Homogeneous Ca--CN--CH--NH Photometry. VII. Metal-Poor Populations in 47~Tucanae (NGC~104)
\footnote{Based on observations made with the Cerro Tololo Inter-American Observatory (CTIO) 1 m telescope, which is operated by the SMARTS consortium.}
\footnote{This work presents results from the European Space Agency (ESA) space mission Gaia. Gaia data are being processed by the Gaia Data Processing and Analysis Consortium (DPAC). Funding for the DPAC is provided by national institutions, in particular the institutions participating in the Gaia MultiLateral Agreement (MLA). The Gaia mission website is \url{https://www.cosmos.esa.int/gaia}. The Gaia archive website is \url{https://archives.esac.esa.int/gaia}.}
}

\author[0000-0002-2122-3030]{Jae-Woo Lee}
\affiliation{Department of Physics and Astronomy, Sejong University\\
209 Neungdong-ro, Gwangjin-Gu, Seoul, 05006, Republic of Korea\\
jaewoolee@sejong.ac.kr, jaewoolee@sejong.edu}

\begin{abstract}
We present new large field-of-view ($\sim$1\arcdeg$\times$1\arcdeg) Ca-CN photometry of the prototypical metal-rich globular cluster 47~Tucanae (NGC~104). Our results are the following.
(1) The populational number ratios of the red giant branch (RGB) and red horizontal branch (RHB) are in excellent agreement: \nrgb\ = 30:70 ($\pm$1--2), where the \cnw\ and \cns\ stand for the CN-weak and CN-strong populations, respectively. Both the \cns\ RGB and RHB populations are more centrally concentrated than those of \cnw\ populations are.
(2) Our photometric metallicities of individual RGB stars in each population can be well described by bimodal distributions with two metallicity peaks, \feh\ $\sim$ $-$0.72 and $-$0.92 dex, where the metal-poor components occupy $\sim$ 13\% of the total RGB stars. The metal-poor populations are more significantly centrally concentrated than the metal-rich populations, showing a similar result that we found in M3.
(3) The RGB bump $V$ magnitudes of individual populations indicate that there is no difference in the helium abundance between the two metal-poor populations, while the helium enhancement of \dy\ $\sim$ 0.02--0.03 is required between the the two metal-rich populations.
(4) The RHB morphology of 47~Tuc appears to support our idea of the bimodal metallicity distribution of the cluster.
We suggest that 47~Tuc could be another example of merger remnants of two globular clusters, similar to M3 and M22.
\end{abstract}

\keywords{Stellar populations (1622); Population II stars (1284); Hertzsprung Russell diagram (725); Globular star clusters (656); Chemical abundances (224); Stellar evolution (1599); Red giant branch (1368); Red giant bump (1369): Horizontal branch stars (746)}

\section{INTRODUCTION}
The ubiquitous nature of the multiple populations (MPs) in globular clusters (GCs) is an enigmatic phenomenon that cannot be easily understood in modern astrophysics \citep[e.g., see][]{bastian18, gratton19, cassisi20}. Once thought to be simple groups of old stars, not only the Galactic but also the extra-Galactic GCs show MPs containing at least two populations: One is called first generation (FG) with chemical compositions similar to abundance patterns observed in Galactic field stars with the same metallicity, while the other is called a second generation (SG) showing the chemical compositions that experienced proton capture processes at high temperature. Most notably, the so-called Na--O and C--N  anticorrelations are thought to be a strong evidence of the MPs in GCs \citep{carretta09, lee10}.

Recent development of numerical simulations may seem promising to delineate the formation of GCs with MPs \citep[e.g.,][]{dercole08, bekki19, calura19, mckenzie21, lacchin21}, however, some fundamental processes that induce the chemical evolution between MPs are yet to be understood \citep[e.g., see][]{bastian15, renzini15, bastian18, gratton19}

47~Tuc (NGC~104) is a massive nearby GC, that has long been considered as a prototypical metal-rich GC \citep[e.g.,][]{hesser87}. Numerous photometric and spectroscopic studies have been conducted for this cluster. Elemental abundance variations in the red giant branch (RGB), red horizontal branch (RHB), and main-sequence (MS) stars in 47~Tuc had been known for decades long before the establishment of the MPs in GCs during the last decade.

\citet{hesser78} employed low-resolution spectroscopy, and he found that there exist CN abundance variations from RGB to MS stars in 47~Tuc \citep[see also][for the abundance variation in MS stars in 47~Tuc]{briley91, briley04, cannon98, harbeck03}. \citet{norris79} confirmed this result, and they concluded that the primordial elemental abundance difference is responsible for such CN anomalies \citep[see also][who drew the same conclusion from the Na--CN positive correlation]{cottrell81}.
The primordial origin of MPs in 47~Tuc was securely confirmed by \citet{carretta04}, who revealed the Na--O anticorrelation in the turn-off and the base of RGB stars in 47~Tuc.
Later, \citet{milone12} nicely demonstrated that the MS split in 47~Tuc can be ascribed to the different helium and light elemental abundances, such as CNO, that can be seen in GC MPs. They also found a radial gradient in the CN abundance, in the sense that the \cns\ RGB stars are more centrally concentrated than the \cnw\ RGB stars are, where the CN weak and strong stars are defined to be stars with weak and strong CN absorption bands at \cnwave\ at a given $V$ magnitude \citep[see also][]{chun79, smith79, briley97, nataf11}.

Not only the RGB stars but also the RHB stars in 47~Tuc also display the prototypical characteristics of MPs. \citet{norris82} found that there exist a CN--CH anticorrelation in the RHB stars in 47~Tuc, and $V$ magnitude of the \cns\ RHB stars is $\Delta V \sim$ 0.04 mag brighter than the \cnw\ counterparts, which led them to conclude that differences in helium abundance or helium core mass are responsible for this luminosity difference between the two groups. \citet{briley97} confirmed this result, and they also noted that the \cns\ RHB stars tend to be bluer than the \cnw\ RHB stars \citep[see also][]{nataf11}. In a similarly context, \citet{gratton13} found that sodium abundances increase toward bluer color along the RHB in 47~Tuc, which can be understood from the existence of the Na--CN positive correlation as we mentioned above.

The presence of the subgiant branch (SGB) split in 47~Tuc also indicates that it is not composed of a simple stellar population. After the monumental discovery of the double SGB sequence in NGC~1851 \citep{milone08,cassisi08}, \citet{anderson09}  reanalyzed a large number of Hubble Space Telescope (HST) archival data, and they found the existence of the double SGB in 47~Tuc, where the second SGB sequence is $\sim$ 0.05 mag fainter than the main SGB body and occupies $\sim$ 10\% of the stars. They also noted that there exists intrinsic color dispersion in the MS stars, which can be ascribed to the metallicity spread (with standard helium abundance) or a helium abundance dispersion of \dy\ = 0.026 \citep[see also][]{milone12}. \citet{dicriscienzo10} interpreted that, in addition to the helium abundance variations in the 47~Tuc, the difference in the C+N+O abundance could be responsible for the double SGB in 47~Tuc, in the sense that the fainter SGB stars contain higher overall C+N+O abundance than the brighter ones. Later, \citet{marino16} performed a spectroscopic study for SGB stars, finding no concrete evidence of a C+N+O enhancement in the faint SGB stars.

Finally, the RGB bump (RGBB) of 47~Tuc also indicates that it is a very intriguing cluster. \citet{nataf11} found that RGBB $V$ magnitude becomes fainter with the radial distance from the cluster's center.

In our previous study, we showed that M3, previously considered as a monometallic prototypical GC, has two groups of stars with different metallicity, \dfeh\ $\sim$ 0.15 dex. More recently, independent studies showed some evidence of a metallicitiy spread in the FG in previously considered monometallic GCs and different metallicity distributions between the FG and SG \citep[e.g.,][]{legnardi22,lardo22}, making the understanding of the formation of GCs a more formidable task.

In this paper, we investigate 47~Tuc using our own photometric system that is optimized to study the GC MPs. So far, almost all the previous studies of the cluster assumed that 47~Tuc shows an unimodal metallicity distribution \citep[see also][]{gratton19}. Interestingly, our results will show that 47~Tuc shows a bimodal metallicity distribution with a metallicity difference of \dfeh\ $\sim$ 0.2 dex, which can be also supported by the RGBB magnitude and the RHB morphology. Our finding will set a crucial constraint to understand the formation of 47~Tuc.

\begin{figure}
\epsscale{1.15}
\figurenum{1}
\plotone{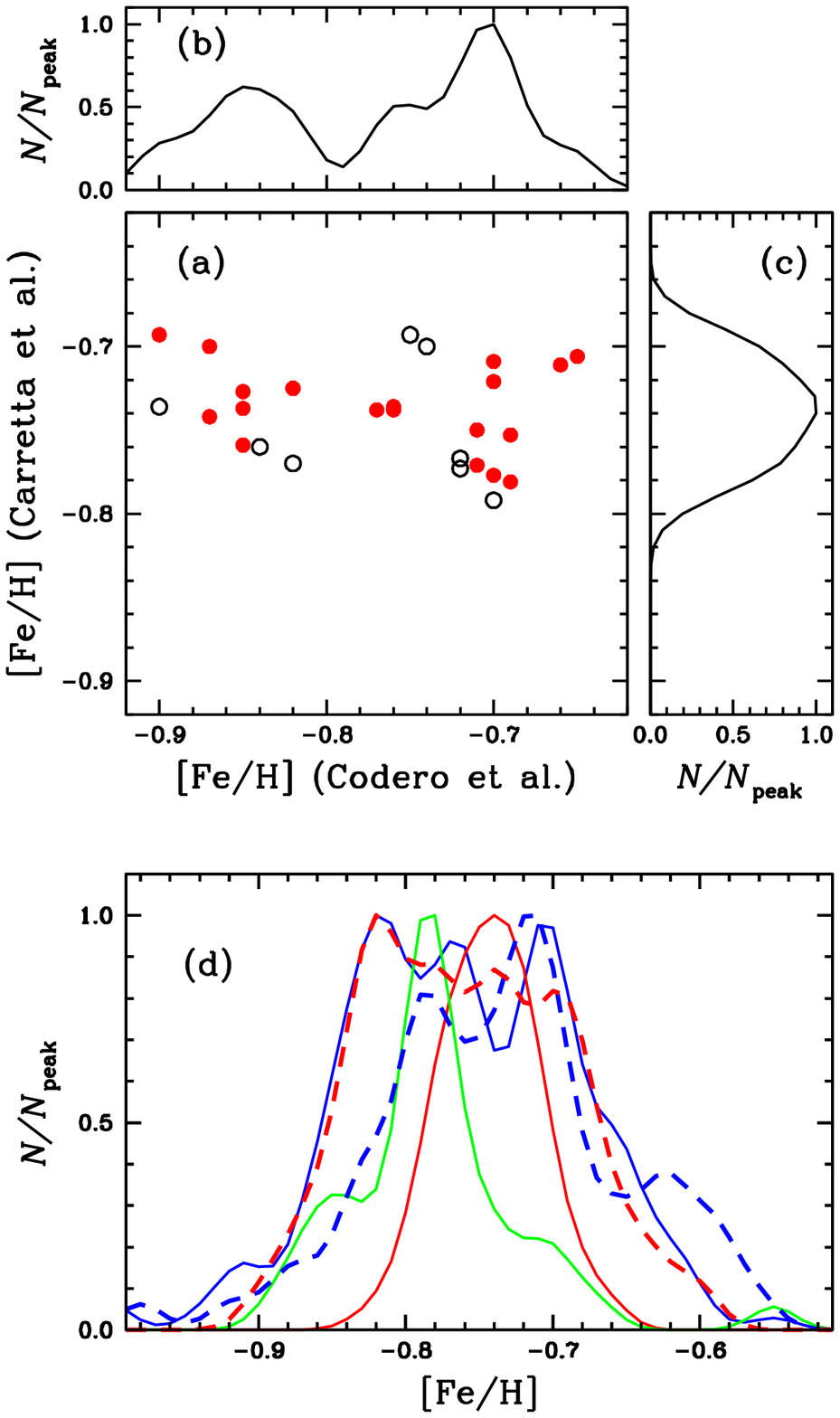}
\caption{
(a) Comparison of \feh\ of \citet{carretta09} against that of \citet{codero14} for 28  RGB stars in 47 Tuc in common between the two studies. The filled red circles indicate the RGB stars using the same instrument between the two studies.
(b) Histogram of the \feh\ distribution by \citet{codero14}.
(c) Histogram of the \feh\ distribution by \citet{carretta09}.
(d) Histograms of the \feh\ measurements from literature. The red, blue, and green solid lines are for the RGB stars by \citet{carretta09}, \citet{codero14}, and \citet{wang17}. The thick red dashed line is for the RHB stars by \citet{gratton13}, while the thick blue dashed line is for the SGB stars by \citet{marino16}. Except for the RGB \feh\ distribution by \citet{carretta09}, other four measurements show nonunimodal \feh\ distributions with large metallicity dispersions.}\label{fig:spec_comp}
\end{figure}

\section{Spectroscopic Metallicity Distributions of 47~Tuc}
Thanks to the advance in high throughput (multiobject) spectrographs combined with large aperture telescopes, utility of high-resolution spectroscopy becomes more important. However, there appears to exist some limitations that significantly reduce the capability of spectroscopic study of GCs. For example, RGB stars in metal-rich GC 47~Tuc show very strong absorption lines combined with severe degree of line blendings that make accurate elemental abundance measurements somewhat difficult \citep[see also][]{lee10, lee16}. Not surprisingly, unequivocal results among different research groups can often be witnessed.

In Figure~\ref{fig:spec_comp}(a), we show a comparison of the \feh\ measurements for 28 RGB stars in 47~Tuc in common with \citet{carretta09} and \citet{codero14}. Using these 28 stars, we obtained the similar mean \feh\ values between the two studies: \mfeh\ = $-$0.739$\pm$0.028$\pm$0.005 dex from \citet{carretta09} and $-$0.769$\pm$0.075$\pm$0.014 dex from \citet{codero14}. However, the \feh\ distributions of individual stars may tell a completely different story as shown in the figure. The \feh\ distribution of \citet{carretta09} appears to be an unimodal, while that of \citet{codero14} is a nonunimodal with a significantly large dispersion. We performed statistical tests to see if these two \feh\ distributions are unimodal. The $p$-value returned from our dip test is 1.85$\times$10$^{-3}$ for \citet{codero14}, indicating that it is nonunimodal. On the other hand, the $p$-value of that of \citet{carretta09} is 0.922, and it is most likely a unimodal, as our histogram suggested. In the figure, the red-filled circles show the RGB stars measured by using the same instrument (FLAMES-GIRAFFE on the Very Large Telescope Kueyen telescope) between the two studies, exhibiting large differences. Therefore, the discrepancy between the two studies is not instrument dependent, but it may have something to do with data analysis procedures in each research group.

In Figure~\ref{fig:spec_comp}(d), we show histograms for \feh\ distributions of individual stars in 47~Tuc from various studies \citep{carretta09, codero14, gratton13, marino16, wang17}. The figure clearly shows that the \feh\ distribution of 47~Tuc is not likely unimodal with large metallicity dispersions, except for that of \citet{carretta09}.

We are left with the impression that \dfeh\ $\sim$ 0.1 -- 0.2 dex from high-resolution spectroscopy can be introduced, which may hamper revealing small-scale structures in the metallicity distribution of given stellar populations \citep[see also][]{lee10, lee16, lee21a}. In this regard, our narrow- and intermediate-band photometry can be a valuable asset to investigate metallicity distributions, for example, of the MPs in GCs.

\begin{figure}
\epsscale{1.15}
\figurenum{2}
\plotone{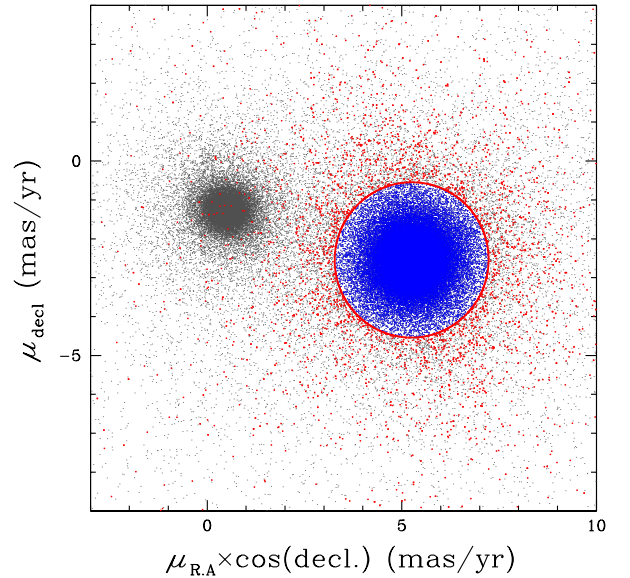}
\caption{
Gaia EDR3 proper-motions of our FOV. The red ellipse indicates the boundary (3$\sigma$) of the 47~Tuc member stars shown with blue dots. The red dots are stars located within 300\arcsec\ from 47~Tuc, and they are classified as nonmember stars based on the proper-motion study.
}\label{fig:pm}
\end{figure}

\begin{figure*}
\epsscale{.85}
\figurenum{3}
\plotone{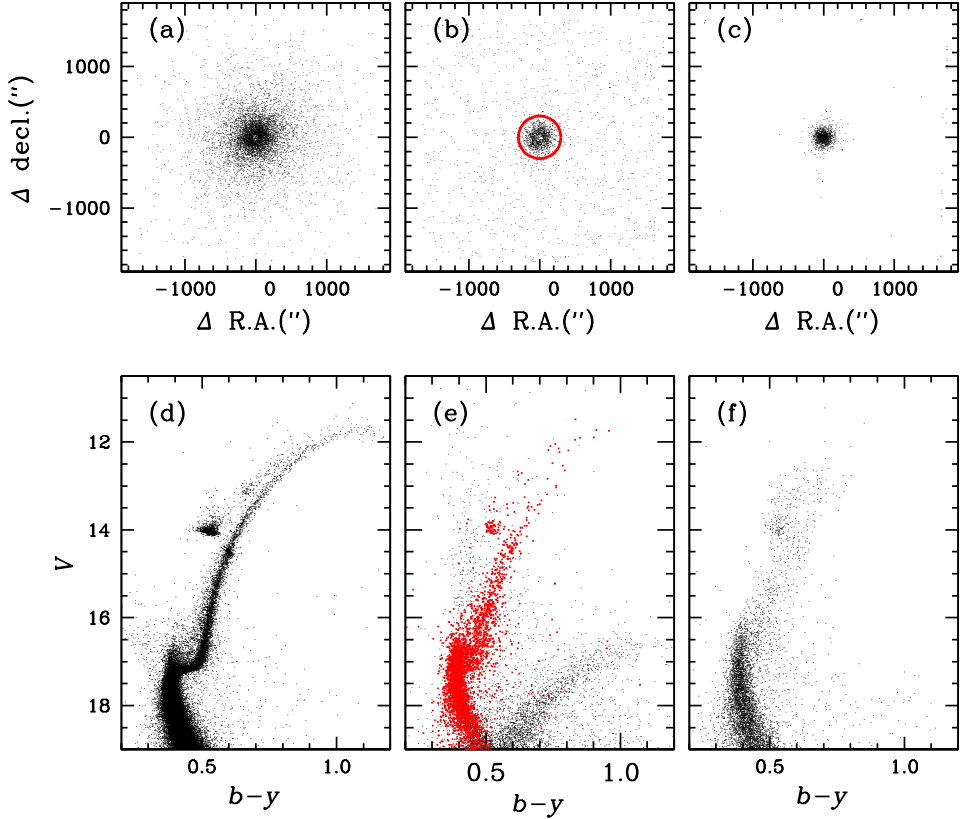}
\caption{
(a) Positions of bright stars with $V \leq$ 17 mag that match with the Gaia EDR3 and classified as 47~Tuc's members.
(b) Positions of bright stars with $V \leq$ 17 mag that match with the Gaia EDR3 but classified as 47~Tuc's nonmembers. The thick red circle indicates the radial distance of 300\arcsec\ from the cluster's center.
(c) Positions of bright stars with $V \leq$ 17 mag that have no counterpart in the Gaia EDR3.
(d) $(b-y)$ versus $V$ CMD for (a).
(e) $(b-y)$ versus $V$ CMD for (b).  Note the presence of the SMC stars in the faint regime. The red dots show the CMD from stars located within 300\arcsec\ from the center that are considered as proper-motion nonmembers of the cluster.
(f) $(b-y)$ versus $V$ CMD for stars with no Gaia EDR3 counterpart.
}\label{fig:cmdpos}
\end{figure*}

\section{OBSERVATIONS AND DATA REDUCTION}\label{s:reduction}
Observations for 47~Tuc were carried out in 54 nights, 27 of which  were photometric, in 9 runs from September 2007 to August 2013 using the CTIO 1.0m telescope. The CTIO 1.0m telescope was equipped with a STA 4k $\times$ 4k CCD camera, providing a plate scale of 0\farcs289 pixel$^{-1}$ and a field-of-view (FOV) of about 20\arcmin\ $\times$ 20\arcmin. The detailed discussion of our new filter system can be found in \citet{lee17}. During the whole seasons, the average seeing of our 47~Tuc science frames was 1\farcs37 $\pm$ 0\farcs20.
The effective FOV of our 47~Tuc science field was about 1\arcdeg\ $\times$ 1\arcdeg\ except for the JWL39 filter. We employed our JWL39 filter only in 2013, and it mostly covered an FOV of 20\arcmin\ $\times$ 20\arcmin\ of the central part of the cluster.

The detailed procedures for the raw data handling were described in our previous works \citep{lee14, lee15, lee17, lp16}. The photometry of the cluster and photometric standard frames were analyzed using DAOPHOTII, DAOGROW, ALLSTAR and ALLFRAME, and associate packages \citep{pbs87, pbs94, lc99}. The total number of stars measured from our ALLFRAME run was about 111,000.

Astrometric solutions for individual stars in our field were derived using the data extracted from the Gaia EDR3 \citep{gaiaedr3} and the IRAF IMCOORS package. Then the astrometric solution was applied to calculate the equatorial coordinates for all stars measured in our science frames.

\begin{figure*}
\epsscale{.85}
\figurenum{4}
\plotone{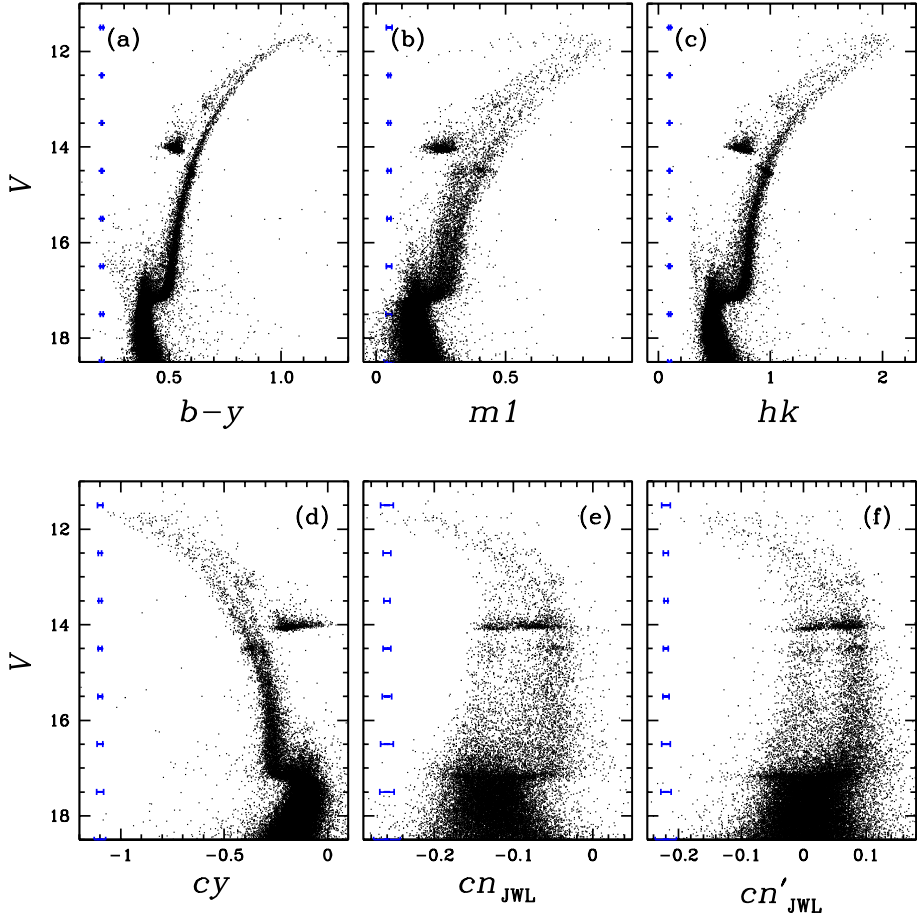}
\caption{CMDs for proper-motion membership stars in 47~Tuc. We also show the mean measurement uncertainties at a given $V$ magnitude bin with blue error bars. Note the broad RGB sequences in $m1$, $cy$ and the discrete double RGB sequences in \cnjwl\ and \cnpjwl\ CMDs.
}\label{fig:cmd}
\end{figure*}

\begin{figure*}
\epsscale{.85}
\figurenum{5}
\plotone{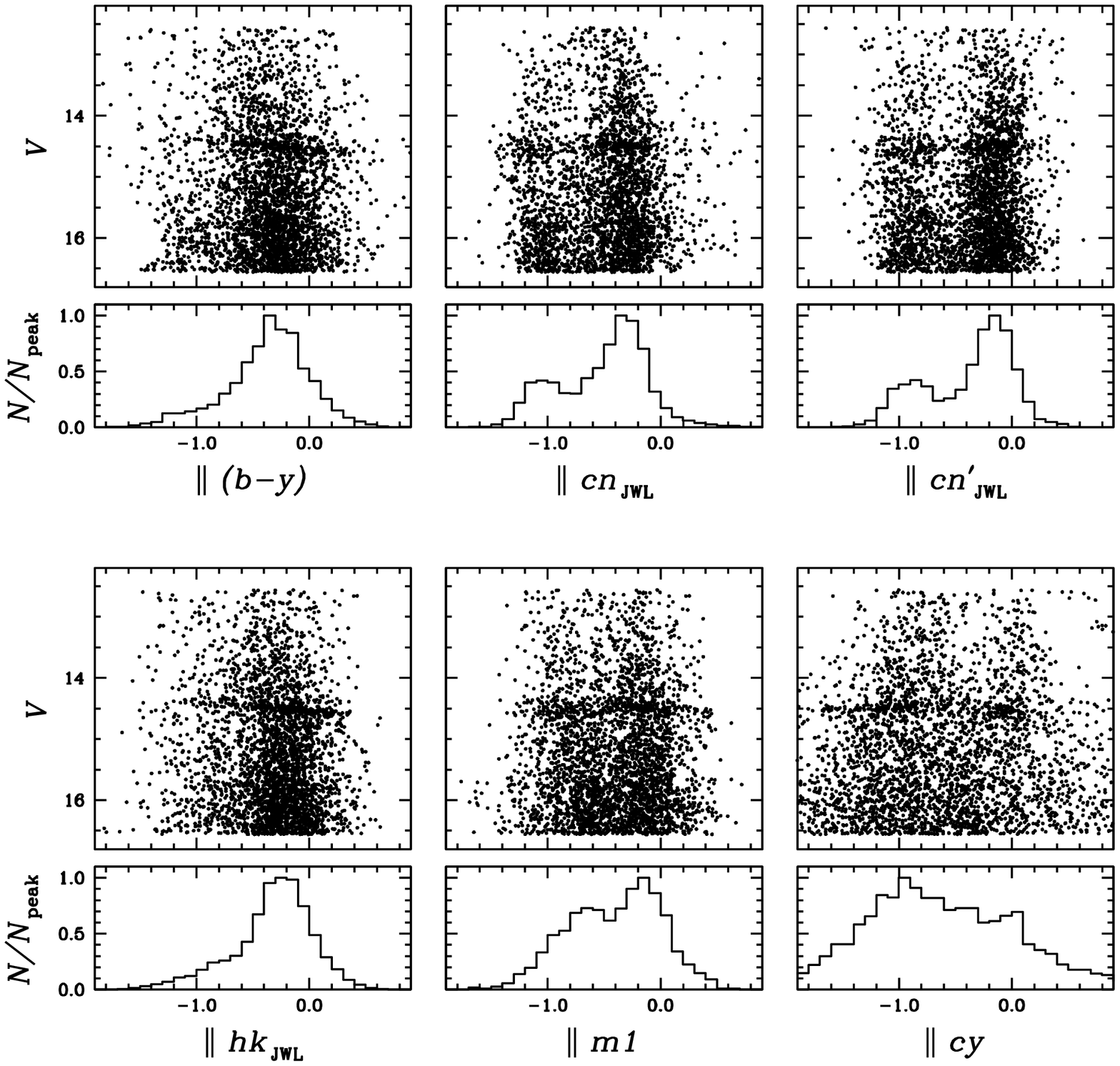}
\caption{
Parallelized CMDs with histograms for various color indices used in our study. Note that our \pcnjwl\ and \pcnpjwl\ indices show discrete double RGB sequences in 47~Tuc. We also note that both the \pby\ and \phkjwl, both of which are good metallicity indicators, show asymmetric distributions toward smaller color indices.
}\label{fig:rec}
\end{figure*}

\section{Membership Selection}\label{s:member}
The Galactic latitude of 47~Tuc is large, $b$ = $-$45\arcdeg, and the contamination from the foreground off-cluster Galactic field stars is expected to be small. On the other hand, due to the proximity to the Small Magellanic Cloud (SMC), the contamination from the SMC can be severe, in particular for the faint regime. In Figure~\ref{fig:pm}, we show the proper-motions for our 47~Tuc's FOV from the Gaia EDR3 \citep{gaiaedr3}, where two groups of stars, one for 47~Tuc and the other for the SMC, can be clearly seen.

In order to select proper-motion member stars, we derived the mean values of proper-motions of 47~Tuc using iterative $\sigma$-clipping calculations, finding that, in units of milliarcsecond per year, ($\mu_{\rm RA}\times\cos\delta$, $\mu_{\rm decl.}$) = (5.253, $-$2.543) with standard deviations along the major axis of the ellipse of 0.666 mas yr$^{-1}$ and along the minor axis of 0.659 mas yr$^{-1}$. We considered that stars within 3$\sigma$ from the mean values are 47~Tuc proper-motion member stars as shown with blue dots in Figure~\ref{fig:pm}.

In Figure~\ref{fig:cmdpos}, we show color-magnitude diagrams (CMDs) for proper-motion members and nonmembers of 47~Tuc. Also shown are the stars without the Gaia EDR3 counterparts. We believe that, due to large measurement uncertainties rising from crowdedness, a significant fraction of 47~Tuc member stars in the central part deviates large in the proper-motion measurements of the Gaia EDR3 as shown in Figure~\ref{fig:pm}. Furthermore, there exists a large number of stars located in the central part of cluster, owing to crowdedness, without proper-motion measurements.
In our current paper, we consider RGB stars with $-$1.5 $\leq$ \vvhb\ $\leq$ 2.5 mag in two cases separately: (1) using proper-motion member stars only (case (1)), where our sample is not complete and tends to be slightly biased toward the outer part of the cluster; (2) using stars selected from the multicolor CMDs (case (2)), which include the stars in case (1) and missing member stars by Gaia EDR3 as shown in Figure~\ref{fig:pm}.

\section{Photometric Indices and Color-Magnitude Diagrams}
We used the filters provided by the CTIO from 2007 to 2010, and we used our own filters since 2012. As we already noted in our previous study \citep{lee15, lee19c}, the original \cactio\ was designed to have a filter bandwidth and pivot wavelength very similar to that of the \caat\ defined by \citet{att91}. Due to aging, the \cactio\ had been degraded, and its original transmission curve had been significantly altered to the shorter wavelength, resulting in having the passband similar to our JWL39 filter \citep[e.g., see Figure~1 of][]{lee19c}. Therefore, we do not use the \cactio\ to measure metallicity. Instead, we used the \cactio\ as a good proxy of the JWL39 filter in our current study.

Throughout this work, we will use our own photometric indices \citep[see also][]{lee19c, lee21a}, defined as
\begin{eqnarray}
hk_{\rm JWL} &=& (Ca_{\rm JWL} - b) - (b-y). \label{eq:hk} \\
cn_{\rm JWL} &=& JWL39 - Ca_{\rm JWL}, \label{eq:cn} \\
cn^\prime_{\rm JWL} &=& Ca_{\rm CTIO} - Ca_{\rm JWL}.  \label{eq:cnp} 
\end{eqnarray}
The \hkjwl\ index is a good photometric measure of metallicity \citep[e.g.,][]{att91, lee09, lee15}, assuming a constant [Ca/Fe] among GC membership stars \citep{carney96, marino19}.  
As we already discussed \citep[][]{lee19c}, the \cnpjwl\ can be used as an excellent proxy of the \cnjwl\ index. Hence, both the \cnjwl\ and \cnpjwl\ indices are photometric measures of the CN band at \cnwave.

In Figure~\ref{fig:cmd}, we show our CMDs. The discrete RGB sequences can be clearly seen both in our \cnjwl\ and \cnpjwl\ CMDs. On the other hand, both the $m1$ and $cy$ CMDs show splits in the upper RGB sequences but not in the lower RGB sequences.

\begin{figure*}
\epsscale{0.85}
\figurenum{6}
\plotone{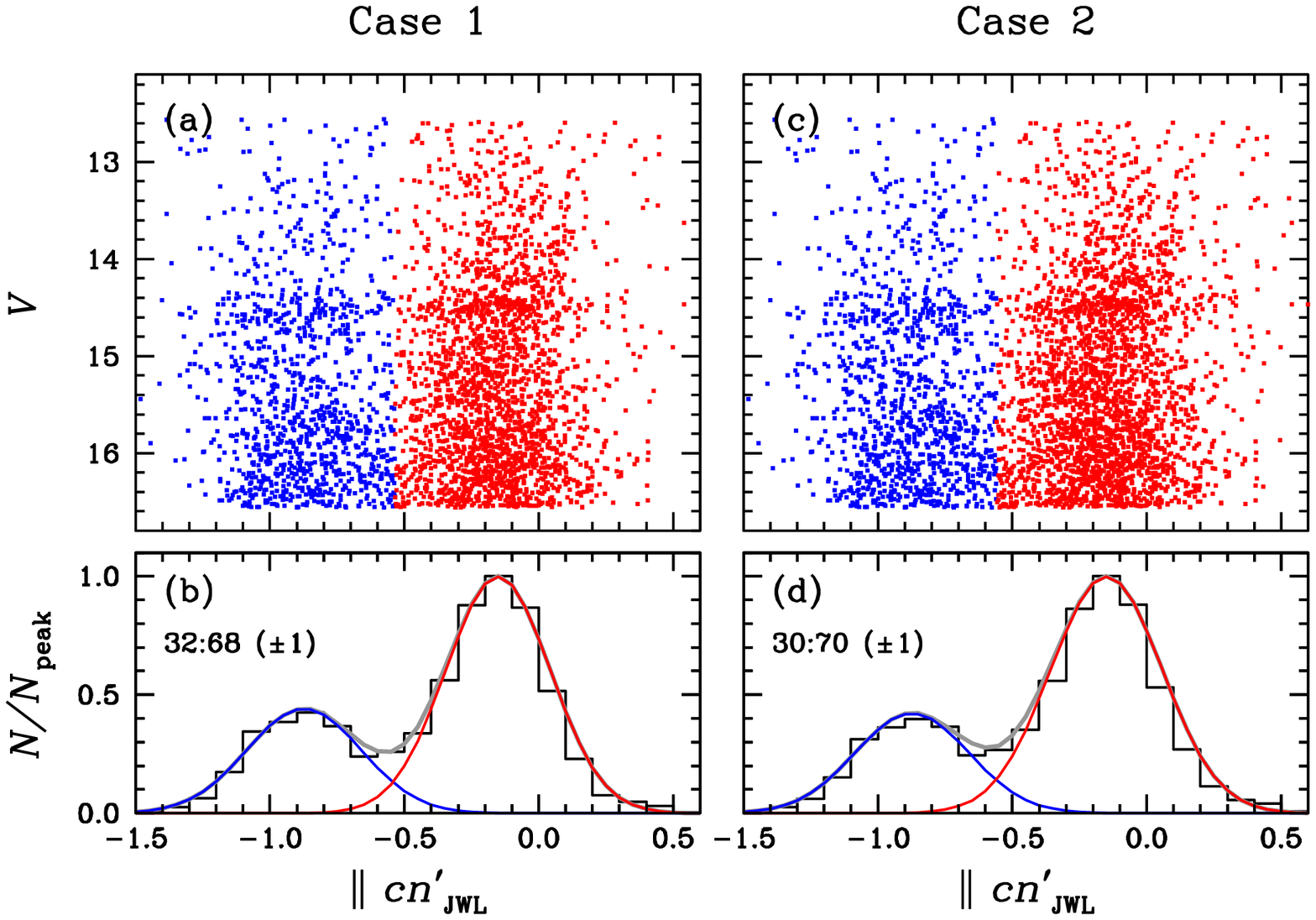}
\caption{
(a) A \pcnpjwl\ vs. $V$ CMD for the proper-motion member RGB stars (case (1)). The blue and red dots denote the \cnw\ and \cns\ RGB stars in 47~Tuc returned from our EM estimator.
(b) The black line denotes the \cnpjwl\ histogram of 47~Tuc RGB stars. The blue and red solid lines are the histograms returned from our EM estimator, and the gray solid line is for that of the total RGB population.
(c) Same as (a) but for all RGB stars (case (2)). (d) Same as (b) but for all RGB stars.
}\label{fig:em}
\end{figure*}

\begin{figure*}
\epsscale{.85}
\figurenum{7}
\plotone{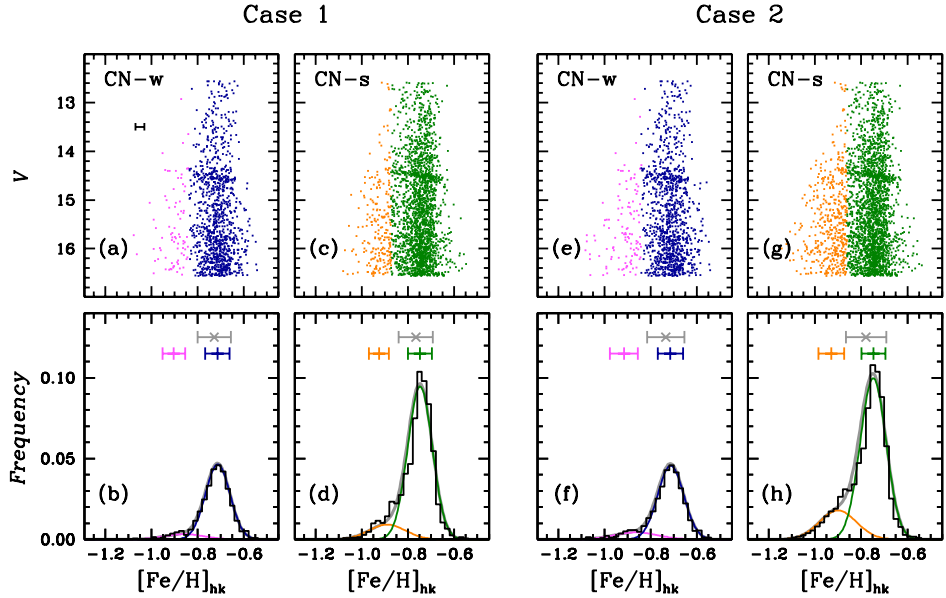}
\caption{
Metallicity distributions of individual populations. The pink, navy, orange, and dark-green colors denote the \cnwp, \cnwr, \cnsp, and \cnsr, respectively.  Gray color denotes the whole population. The error bars are for $\pm1\sigma$.
In upper left corner of the panel (a), we show the measurement uncertainties ($\pm1\sigma$) using the bootstrap method, $\pm$0.019 dex.
}\label{fig:feh}
\end{figure*}

\section{Red Giant Branch}\label{s:rgb}

\subsection{Populational Tagging}\label{ss:tag}
We followed the similar method that we developed in our previous studies of other GCs with MPs \citep[e.g., see][]{lee21a, lee21b}.
In order to remove the luminosity effect in our analysis, the RGB sequences in the individual color indices were parallelized using the following relation \citep[also see][]{lee19a, lee19c};
\begin{equation}
\parallel{\rm CI}(x) \equiv \frac{{\rm CI}(x) - {\rm CI}_{\rm red}}
{{\rm CI}_{\rm red}-{\rm CI}_{\rm blue}},\label{eq1}\label{eq:pl}
\end{equation}
where CI$(x)$ is the color index of the individual stars, and CI$_{\rm red}$, CI$_{\rm blue}$ are color indices for the fiducials of the red and the blue sequences of individual color indices, respectively \citep[see also][]{milone17}.
We derived fourth-order polynomial fits for individual color indices, and we show our parallelized CMDs and histograms in Figure~\ref{fig:rec}. Note that our \pcnjwl\ and \pcnpjwl\ show discrete bimodal RGB distributions, a strong evidence of MPs in 47~Tuc. As we showed in Figure~\ref{fig:cmd}, our measurement uncertainties are significantly smaller than the widths of RGB sequences in individual color indices. Therefore, double RGB sequences in \pcnjwl\ and \pcnpjwl\ are thought to be real, reflecting a bimodal CN distribution of RGB stars in 47~Tuc  \citep[e.g.,][]{hesser78, norris79, briley91, lee19b}.
The \pmone\ and \pcy\ CMDs also suggest MPs in 47~Tuc, but the populational separations in the lower part of the RGB sequence are somewhat ambiguous as we already pointed out.
We also note asymmetric distributions of RGB stars in the \pby\ and \phkjwl, extending toward smaller color index values, most likely due to the difference in metallicity in RGB stars in 47~Tuc as we will discuss below.

In our previous studies \citep{lee17, lee18, lee19a, lee19c, lee20, lee21a}, we showed that both the \cnjwl\ and \cnpjwl\ indices are very powerful tools to classify MPs in GCs.
As we already mentioned, our \cnjwl\ observations were limited to the central part of the cluster, and we decided to use the \cnpjwl\ to classify MPs in 47~Tuc in our current study.
Assuming a bimodal \pcnpjwl\ distribution as already shown in Figure~\ref{fig:rec}, we applied the expectation maximization (EM) algorithm for a two-component Gaussian mixture model on our \pcnpjwl\ index in order to perform a populational tagging of RGB stars in 47~Tuc.
The stars with $P$(\pcnpjwl$|x_i) \geq$ 0.5 from the EM estimator correspond to the \cnw\ population, where $x_i$ denotes the individual RGB stars, while those with $P$(\pcnpjwl$|x_i)$ $<$ 0.5  correspond to the \cns\ population. In an iterative manner, we calculated the probability of individual stars for being the \cnw\ and the \cns\ populations.
Through this process, we obtained the number ratio between the two populations \nrgb\ = 32:68 ($\pm$1) for the proper-motion member RGB stars (case (1)) and 30:70 ($\pm$1) for all RGB stars (case (2)). We show our result in Figure~\ref{fig:em}.
As we will show later, the \cns\ population is more centrally concentrated than the \cnw\ is, and the incomplete detection is more severe for the \cns\ stars in the central part of cluster, resulting in a slightly smaller \cns\ fraction when the proper-motion-selected RGB stars were used.
Note that our result is significantly different from that of \citet{milone17}. They relied on the HST photometry with a small FOV, $\sim$ 3\arcmin$\times$3\arcmin, compared to our observation, $\sim$ 1\arcdeg$\times$1\arcdeg, and they obtained $N_1/N_{\rm tot}$ = 0.175$\pm$0.009, where the $N_1$ denotes the number of the FG of stars.

We emphasize that the FG of stars defined by \citet{milone17} is corresponding to our \cnw\ population \citep[see][]{lee17}.
The discrepancy between our result and that of \citet{milone17} is a natural consequence of a small FOV of the HST observations of \citet{milone17} with a strong populational radial gradient in 47~Tuc. We note that \citet{milone12} obtained $\sim$ 30\% of the FG of stars by combining the ground-based observations with a large FOV, consistent with our results.

\begin{deluxetable}{cll}[t]
\tablenum{1}
\tablecaption{Abundances for input model atmospheres.\label{tab:input}}
\tablewidth{0pc}
\tablehead{
\multicolumn{1}{c}{Element} &
\multicolumn{1}{c}{Abundance} &
\multicolumn{1}{c}{Note}
}
\startdata
\feh    &  $-$0.5, $-$0.7, $-$0.9 & \\
$Y$       &  0.25, 0.28  & \\
([C, N, O/Fe])  & (0.00, 0.00, 0.30), ($-$0.30, 1.00, 0.00) & CNO1\\
      & (0.15, 0.15, 0.50), ($-$0.45, 1.35, 0.10) & CNO2\\
\enddata
\end{deluxetable}

\begin{deluxetable*}{lcccr}[t]
\tablenum{2}
\tablecaption{Mean \feh\ values.\label{tab:feh}}
\tablewidth{0pc}
\tablehead{
\multicolumn{1}{c}{} &
\multicolumn{1}{c}{Dartmouth} &
\multicolumn{1}{c}{$Y^2$} &
\multicolumn{1}{c}{PGPUC} &
\multicolumn{1}{c}{Freq.} 
\\
\multicolumn{1}{c}{} &
\multicolumn{1}{c}{$\langle$\fehhk$\rangle$} &
\multicolumn{1}{c}{$\langle$\fehhk$\rangle$} &
\multicolumn{1}{c}{$\langle$\fehhk$\rangle$} &
\multicolumn{1}{c}{(\%)} 
}
\startdata
\multicolumn{5}{c}{Case (1): Proper-motion member (Gaia EDR3)}\\
\hline
Mean  & $-$0.753$\pm$0.075$\pm$0.001 & $-$0.734$\pm$0.074$\pm$0.001 & $-$0.730$\pm$0.074$\pm$0.001 & \\
& & & & \\
\cnw  & $-$0.728$\pm$0.072$\pm$0.002 & $-$0.731$\pm$0.073$\pm$0.002 & $-$0.719$\pm$0.073$\pm$0.002 & 32.3 \\
\cnwp & $-$0.903$\pm$0.049$\pm$0.005 & $-$0.900$\pm$0.046$\pm$0.005 & $-$0.901$\pm$0.049$\pm$0.006 &  2.3 \\
\cnwr & $-$0.714$\pm$0.053$\pm$0.002 & $-$0.717$\pm$0.055$\pm$0.002 & $-$0.705$\pm$0.053$\pm$0.002 & 30.0 \\
& & & & \\
\cns  & $-$0.766$\pm$0.074$\pm$0.002 & $-$0.736$\pm$0.074$\pm$0.002 & $-$0.735$\pm$0.075$\pm$0.002 & 67.7 \\
\cnsp & $-$0.936$\pm$0.044$\pm$0.003 & $-$0.911$\pm$0.037$\pm$0.003 & $-$0.914$\pm$0.040$\pm$0.003 &  6.6 \\
\cnsr & $-$0.748$\pm$0.052$\pm$0.001 & $-$0.723$\pm$0.058$\pm$0.001 & $-$0.722$\pm$0.057$\pm$0.001 & 61.1 \\
\hline\hline
\multicolumn{5}{c}{Case (2): All RGBs}\\
\hline
Mean  & $-$0.766$\pm$0.088$\pm$0.001 & $-$0.746$\pm$0.085$\pm$0.001 & $-$0.742$\pm$0.087$\pm$0.001 & \\
& & & & \\
\cnw  & $-$0.735$\pm$0.081$\pm$0.002 & $-$0.738$\pm$0.082$\pm$0.002 & $-$0.726$\pm$0.083$\pm$0.002 & 29.7 \\
\cnwp & $-$0.916$\pm$0.060$\pm$0.006 & $-$0.911$\pm$0.055$\pm$0.005 & $-$0.915$\pm$0.060$\pm$0.006 &  2.8 \\
\cnwr & $-$0.715$\pm$0.055$\pm$0.002 & $-$0.717$\pm$0.057$\pm$0.002 & $-$0.706$\pm$0.056$\pm$0.002 & 26.9 \\
& & & & \\
\cns  & $-$0.779$\pm$0.087$\pm$0.002 & $-$0.749$\pm$0.086$\pm$0.002 & $-$0.749$\pm$0.088$\pm$0.002 & 70.3 \\
\cnsp & $-$0.929$\pm$0.055$\pm$0.003 & $-$0.913$\pm$0.048$\pm$0.003 & $-$0.916$\pm$0.052$\pm$0.003 &  9.8 \\
\cnsr & $-$0.747$\pm$0.052$\pm$0.001 & $-$0.724$\pm$0.058$\pm$0.001 & $-$0.722$\pm$0.057$\pm$0.001 & 60.5 \\
\enddata
\end{deluxetable*}

\begin{deluxetable}{lr}[t]
\tablenum{3}
\tablecaption{Differences in Mean \feh\ Values Using Different Isochrones.\label{tab:dfeh}}
\tablewidth{0pc}
\tablehead{
\multicolumn{1}{c}{} &
\multicolumn{1}{c}{$\Delta\langle$[Fe/H]$\rangle$}
}
\startdata
$\Delta$\fehhk(Dartmouth $-$ $Y^2$) &    0.003$\pm$0.008$\pm$0.000 \\
$\Delta$\fehhk(Dartmouth $-$ PGPUC) & $-$0.020$\pm$0.012$\pm$0.000 \\
$\Delta$\fehhk($Y^2$ $-$ PGPUC)     & $-$0.006$\pm$0.008$\pm$0.000 \\
\enddata
\end{deluxetable}

\begin{deluxetable*}{lccc}[t]
\tablenum{4}
\tablecaption{Differences in Mean \feh\ Values Using Different CNO Abundances.\label{tab:dcno}}
\tablewidth{0pc}
\tablehead{
\multicolumn{1}{c}{} &
\multicolumn{3}{c}{$\langle$[Fe/H]$_{\rm CNO1} - $[Fe/H]$_{\rm CNO2}\rangle$}\\
\cline{2-4}
\multicolumn{1}{c}{} &
\multicolumn{1}{c}{All} &
\multicolumn{1}{c}{\cnw} &
\multicolumn{1}{c}{\cns}
}
\startdata
Dartmouth &    $-$0.003$\pm$0.007$\pm$0.000 & 0.006$\pm$0.002$\pm$0.000 & $-$0.007$\pm$0.008$\pm$0.000 \\
PGPUC &    $-$0.003$\pm$0.006$\pm$0.000 & 0.006$\pm$0.002$\pm$0.000 & $-$0.007$\pm$0.008$\pm$0.000 \\
YAPSI     &    $-$0.003$\pm$0.007$\pm$0.000 & 0.007$\pm$0.001$\pm$0.000 & $-$0.008$\pm$0.009$\pm$0.000 \\
\enddata
\end{deluxetable*}

\begin{figure*}
\epsscale{.85}
\figurenum{8}
\plotone{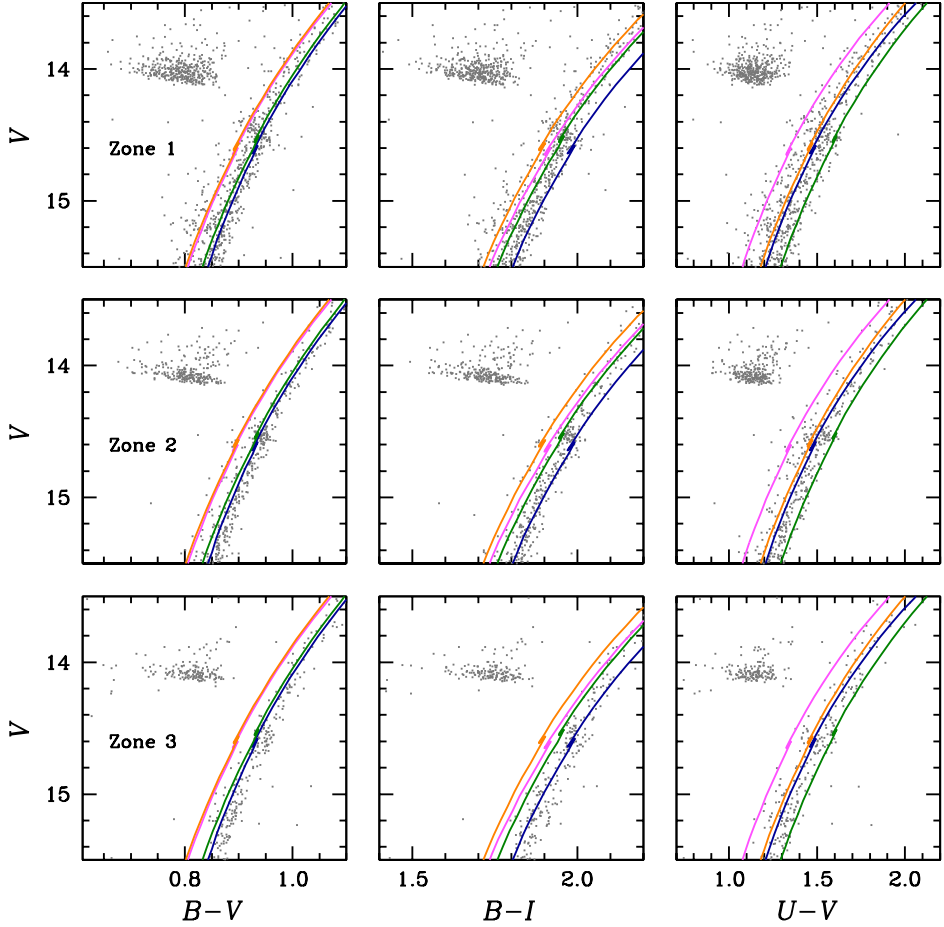}
\caption{
The $(B-V)$, $(B-I)$, and $(U-V)$ CMDs using the multi-color photometry of \citet{pbs19} with three different radial zones, 60\arcsec\ $\leq$ $r$ $<$ 200\arcsec\ (zone (1)), 200\arcsec\ $\leq$ $r$ $<$ 400\arcsec\ (zone (2)), and 400\arcsec\ $\leq$ $r$ $<$ 600\arcsec\ (zone (3)), in 47~Tuc. Note that we used stars with low photometric measurement uncertainties, $\sigma(B-V) \leq 0.01$ mag.
The pink, navy, orange, and dark-green colors denote model isochrones with (\feh, $Y$, CNO) = ($-$0.90 dex, 0.25, \cnw), ($-$0.70 dex, 0.25, \cnw), ($-$0.90 dex, 0.25, \cns), and ($-$0.70 dex, 0.28, \cns), respectively, for 12 Gyr.
}\label{fig:stetson}
\end{figure*}

\begin{deluxetable}{lcc}[t]
\tablenum{5}
\tablecaption{RGBB $V$ magnitudes.\label{tab:rgbbv}}
\tablewidth{0pc}
\tablehead{
\multicolumn{1}{c}{Pop.} &
\multicolumn{1}{c}{This Study} &
\multicolumn{1}{c}{\citet{pbs19}}
}
\startdata
\cnwp & 14.399 ($\pm$0.040) & 14.408 ($\pm$0.040) \\
\cnwr & 14.559 ($\pm$0.025) & 14.530 ($\pm$0.025) \\
\cnsp & 14.385 ($\pm$0.035) & 14.358 ($\pm$0.030) \\
\cnsr & 14.492 ($\pm$0.020) & 14.525 ($\pm$0.020) \\
\enddata
\end{deluxetable}

\subsection{Metallicity Distributions}
As we mentioned earlier, our \hkjwl\ index is an excellent measure of the \ion{Ca}{2} H \& K lines \citep[see][and references therein]{att91, lee09, lee15}. Providing a constant [Ca/Fe] value in a given GC \citep[e.g.,][]{carney96, marino19}, the \hkjwl\ index can be a good measure of metallicity. In our previous studies, we obtained the \feh\ values of individual RGB stars in M3 and M5 using our \hkjwl\ \citep{lee21a, lee21b}. Here, we applied same procedure to derive metallicities of individual RGB stars in 47~Tuc.

To derive photometric metallicity of individual RGB stars, we used three different model isochrones, the Dartmouth \citep{dartmouth}, PGPUC \citep{valcarce12}, and $Y^2$ \citep{y2}.
We obtained model grids for \feh\ = $-$0.9, $-$0.7, and $-$0.5 dex with [$\alpha$/Fe] = +0.3 dex, $Y$ = 0.25 and 0.28, and the age of 12 Gyr \citep[e.g., see][]{brogaard17}.
We constructed series of model atmospheres and synthetic spectra using ATLAS12 and SYNTHE \citep{kurucz11}. During our calculations, we adopted two sets of CNO abundances between the \cnw\ and \cns\ populations to explore how the CNO abundances may affect our results. For one set, we adopted ([C/Fe], [N/Fe], [O/Fe]) of (0.00, 0.00, 0.30) for the \cnw\ and ($-$0.30, 1.00, 0.00) for the \cns\ (CNO1, see Table~\ref{tab:input}). For the other, we adopted (0.15, 0.15, 0.50) for the \cnw\ and ($-$0.45, 1.35, 0.10) for the \cns\ (CNO2).
Individual synthetic spectra were convolved with our filter transmission functions to be converted to our photometric system.
Finally we calculated bolometric corrections for individual colors using the same  method described by \citet{giradi02}.

The photometric metallicity of individual RGB stars can be calculated using the following relation \citep[e.g., see][]{lee21a};
\begin{eqnarray}
{\rm [Fe/H]}_{hk} &\approx& f(hk_{\rm JWL},~ M_V),\label{eq:FeH}
\end{eqnarray}
and we show our results in Table~\ref{tab:feh}.
We obtained the mean photometric metallicity of $\langle$\fehhk$\rangle$ $\approx$ $-$0.75 dex for 47~Tuc, and our result is consistent with that of previous results by others \citep[e.g., see][]{carretta09, codero14, marino16} to within measurement uncertainties.
Not surprisingly, the metallicity from the proper-motion member RGB stars (case (1)) and that from all RGB stars (case (2)) are in excellent agreement.
As shown in Tables~\ref{tab:feh} and \ref{tab:dfeh}, the differences in the photometric metallicity returned from different model isochrones are too small to affect our results, $\Delta$\feh\ $\leq$ 0.02 dex.
We would like to emphasize that we are interested in the relative \feh\ values among RGB stars, and the differences in the relative \feh\ distributions from different model isochrones are negligibly small.
Also importantly, the differences in metallicity from different CNO abundances are also negligibly small. In Table~\ref{tab:dcno}, we show \dfeh\ between those returned from the CNO1 and CNO2 compositions. We obtained the metallicity difference of \dfeh\ $\leq$ 0.01 dex, which may imply that the metallicity spread in 47~Tuc RGB stars described below may not be related to our adopted CNO abundances of individual populations.
In the following analysis, we will take the \fehhk\ returned from the Dartmouth isochrones as our reference photometric metallicity in order to maintain consistency with our previous works on M3 and M5 \citep{lee21a, lee21b}.
We emphasize again that the choice of our reference values does not affect our results.

In Figure~\ref{fig:feh}, we show the metallicity distributions for the \cnw\ and \cns\ populations. As already shown, there is no metallicity difference between the \cnw\ and \cns\ populations, same as in our previous results for M3 and M5 \citep{lee21a,lee21b}.
On the other hand, both the \cnw\ and \cns\ populations exhibit asymmetric metallicity distributions with long metal-poor tails, suggestive of multimodal metallicity distributions.
We estimated measurement uncertainties using the bootstrap method, and we obtained $\pm$0.019 dex for both populations. As shown in Figure~\ref{fig:feh}, our estimated measurement error is too small to fully explain the metallicity spread seen in each population, and we believe that the intrinsic metallicity differences among constituent stars are responsible for the asymmetric metallicity distributions.
In order to examine multimodal metallicity distributions of 47~Tuc,
we applied an EM analysis with a two-component Gaussian mixture model for the metallicity distributions of both the \cnw\ and \cns\ populations.
We show our results in Table~\ref{tab:feh} and Figure~\ref{fig:feh}, suggesting that each population can be well described by bimodal metallicity distributions, reminiscent of M3 \citep{lee21a}. The metal poor components (the \cnwp\ and \cnsp), which constitute about 10 -- 13\% of the total RGB population, are about 0.2 dex more metal-poor than the metal-rich components (the \cnwr\ and \cnsr).

The $UBVI$ photometry of \citet{pbs19} appears to support our results that 47~Tuc contains metal-poor components. In Figure~\ref{fig:stetson}, we show the $(B-V)$, $(B-I)$, and $(U-V)$ CMDs with three different radial zones, 60\arcsec\ $\leq$ $r$ $<$ 200\arcsec\ (zone (1)), 200\arcsec\ $\leq$ $r$ $<$ 400\arcsec\ (zone (2)), and 400\arcsec\ $\leq$ $r$ $<$ 600\arcsec\ (zone (3)), in 47~Tuc. Note that only the $V$ passband is not affected by the carbon and nitrogen abundance variations. Other photometric passbands, $U$, $B$, and $I$ contain CN and CH molecular bands, and therefore, their magnitudes can be altered depending on their CNO abundances (i.e., the \cnw\ and \cns). Interestingly, $B$ passband contains both the CN and CH bands \citep[see Figure~1 of][]{lee19c}.
In GC stars, where the carbon and nitrogen abundances are anticorrelated, the contribution of the CN molecules tends to be canceled out by that of the CH molecules \citep[see also][]{sbordone11} . Therefore, $(B-V)$ color is almost unaffected by the carbon and nitrogen abundance variations. In Figure~8, one can find that the $(B-V)$ colors of the model isochrone for the \cnwp\ are almost identical to those of the \cnsp. The same is true for the \cnwr\ and \cnsr.
The figure suggests that 47~Tuc contains metal-poor population in the central part of the cluster with a radial gradient, supporting our results discussed above.

\begin{figure*}
\epsscale{.85}
\figurenum{9}
\plotone{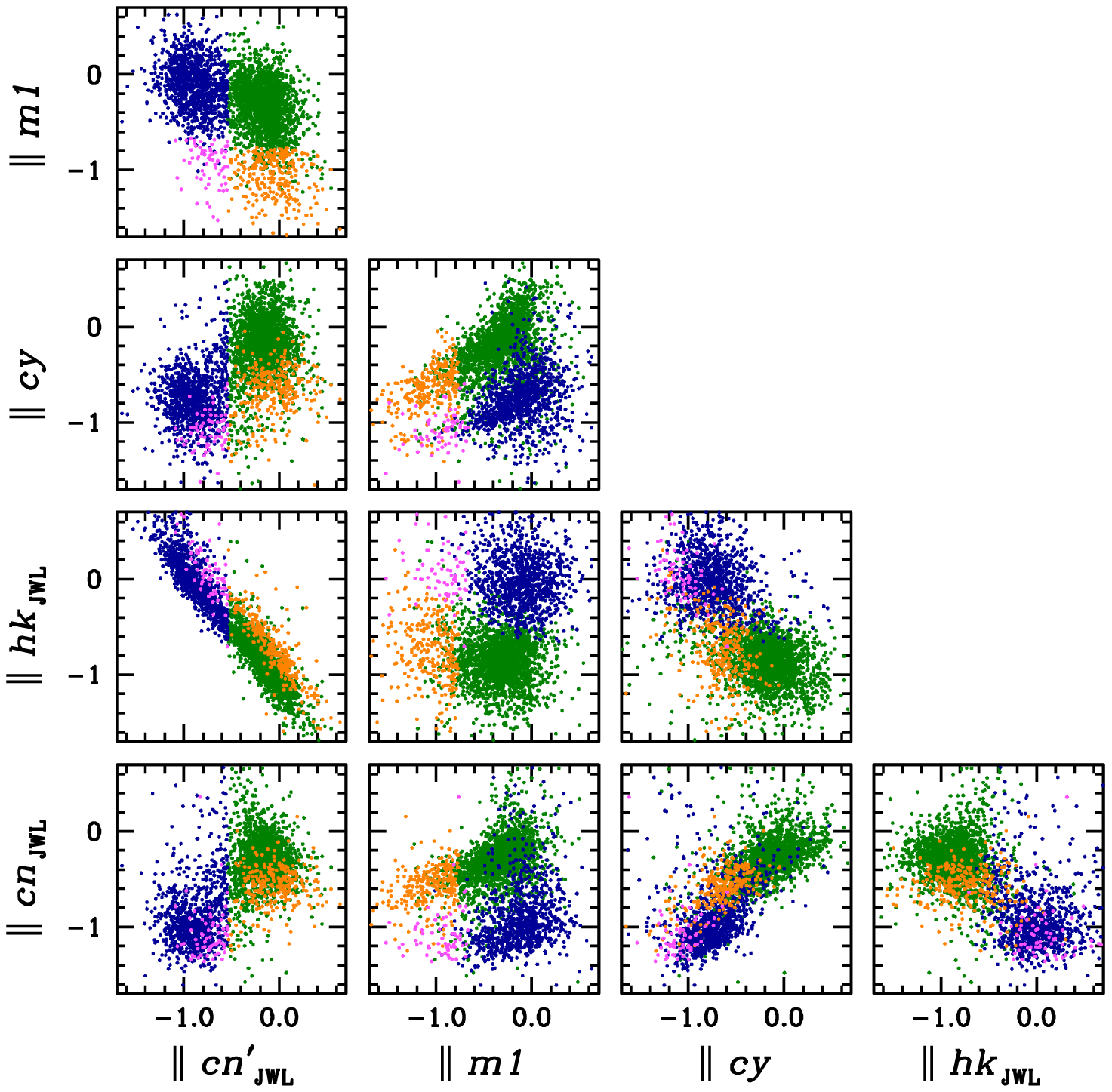}
\caption{
Comparisons of parallelized color indices of the proper-motion membership 47~Tuc RGB stars with $-$1.5 $\leq$ \vvhb\ $\leq$ 2.5 mag, showing that our four groups of stars are different stellar populations. Colors are the same as Figure~\ref{fig:feh}.
}\label{fig:ci}
\end{figure*}

In Figure~\ref{fig:ci}, we show comparisons of individual parallelized color indices of the proper-motion membership RGB stars in 47~Tuc. The figure clearly suggests that the four groups of stars that we obtained are truly  separate populations occupying their own characteristic domains.

\begin{figure*}
\epsscale{.85}
\figurenum{10}
\plotone{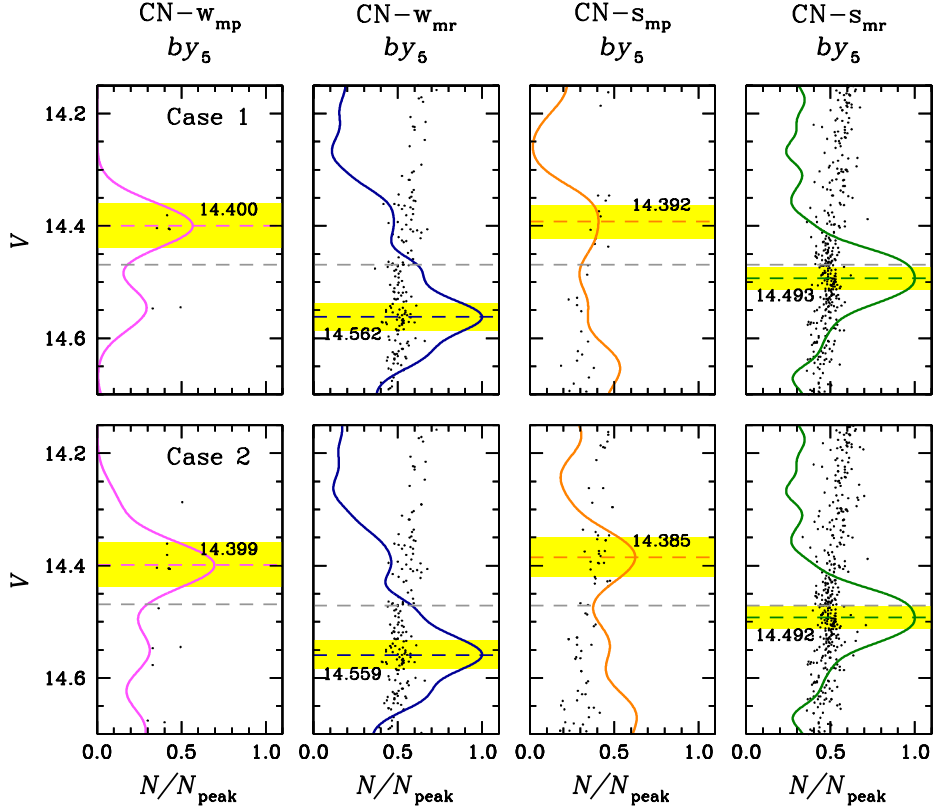}
\caption{
Plots of $by_5$ vs. $V$ CMDs and differential LFs of individual populations. We show the mean RGBB $V$ magnitudes of individual populations with gray dotted lines and those of individual populations with dotted lines with respective colors and yellow-shaded boxes. Note that the $by_5$ values are defined to be $by_5 = 5\times(b-y) - 2.5$ for the clarity of the figure.
}\label{fig:bump}
\end{figure*}

\begin{figure}
\epsscale{1.2}
\figurenum{11}
\plotone{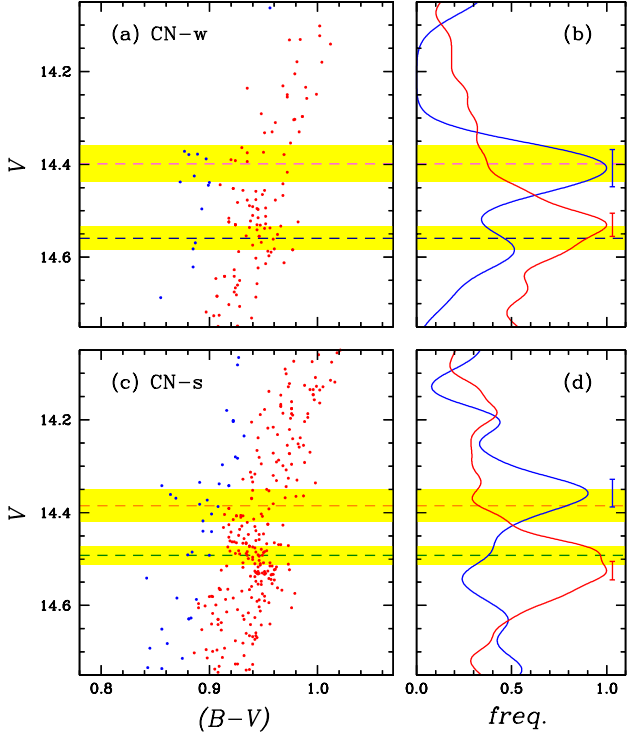}
\caption{
(a) Plots of $(B-V)$ vs. $V$ CMD for the \cnw\ population using the photometric data by \citet{pbs19}. The blue and red dots denote the \cnwp\ and \cnwr\ stars based on the $\parallel (B-V)$ distribution. The horizontal long-dashed lines and yellow boxes denote the \vbump\ and measurement uncertainties for the \cnwp\ (pink) and \cnwr\ (navy) populations based on our own photometry.
(b) Differential LFs of the \cnwp\ (blue) and \cnwr\ (red). The error bars denote the measurement uncertainties using the data by \citet{pbs19}.
(c) Same as (a) but for the \cns\ population.
(d) Same as (b) but for the \cns\ population.
}\label{fig:bump_stetson}
\end{figure}

\begin{figure*}
\epsscale{0.85}
\figurenum{12}
\plotone{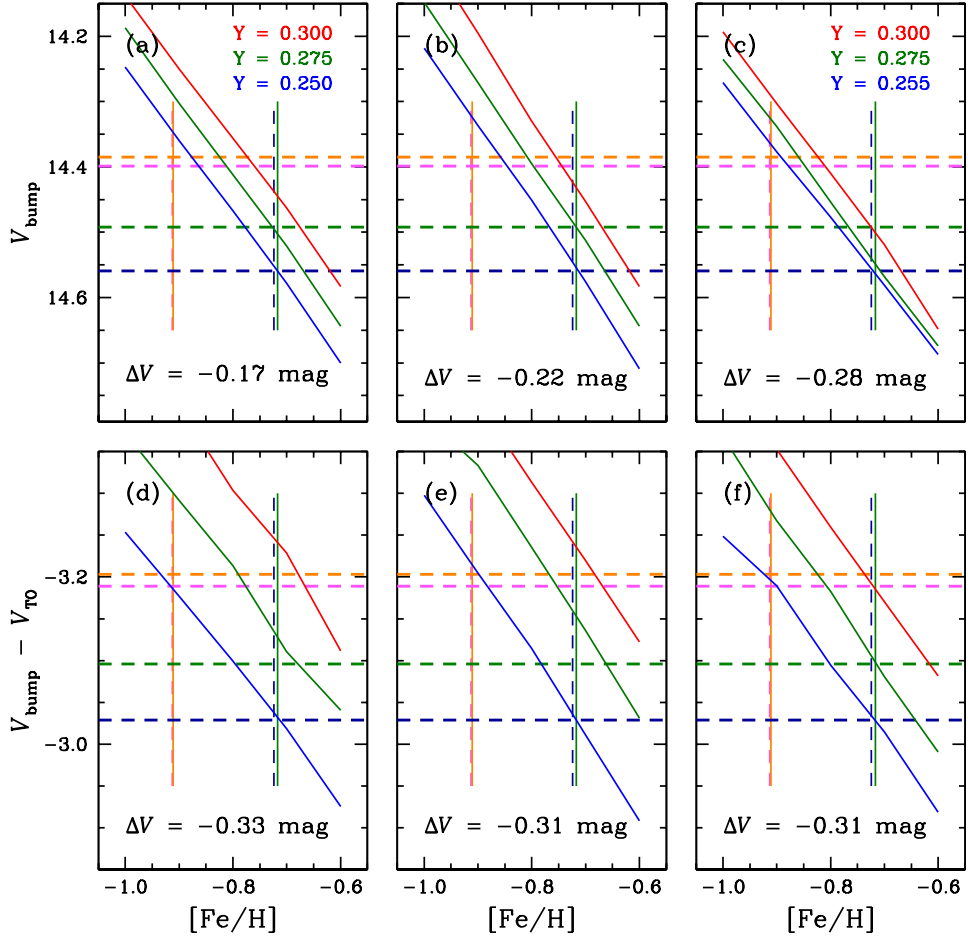}
\caption{
(a) A plot of \vbump\ versus \feh\ using the Y$^2$ isochrones for $Y$ = 0.250, 0.275, and 0.300 along with those of four populations in 47~Tuc. The $\Delta V$ indicates the $V$ magnitude offset in model isochrones to match our \vbump\ magnitude the \cnwr\ population. The colors are the same as Figure~\ref{fig:feh}.
(b) Same as (a) but using the PGPUC isochrones.
(c) A plot of \vbump\ versus \feh\ using the BaSTI isochrones for $Y$ = 0.255, 0.275, and 0.300.
(d) A plot of (\vbumpto) versus \feh\ using the $Y^2$ isochrones.
(e) Same as (d) but using the PGPUC isochrones.
(f) Same as (d) but using the BaSTI isochrones.
}\label{fig:synbump}
\end{figure*}

\begin{deluxetable*}{lccc}[t]
\tablenum{6}
\tablecaption{Slopes in \vbump\ magnitudes versus \feh\ and $Y$.\label{tab:rgbbslope}}
\tablewidth{0pc}
\tablehead{
\multicolumn{1}{c}{Relation} &
\multicolumn{3}{c}{Slope $a$}\\
\cline{2-4}
\multicolumn{1}{c}{($\Delta$ Abundance)} &
\multicolumn{1}{c}{$Y^2$} &
\multicolumn{1}{c}{PGPUC} &
\multicolumn{1}{c}{BaSTI}}
\startdata
(1) \vbump\ $\propto$ $a\times Y$ & $-$2.279$\pm$0.028 & $-$2.425$\pm$0.033 & $-$1.571$\pm$0.134 \\
\dy\tablenotemark{1} & 0.029$\pm$0.014 & 0.028$\pm$0.013 & 0.043$\pm$0.021 \\
 & & & \\
(2) (\vbump\ $- V_{\rm TO}$) $\propto$ $a\times Y$ & $-$4.267$\pm$0.163 & $-$3.947$\pm$0.192 & $-$3.790$\pm$0.504 \\
\dy\tablenotemark{1} & 0.016$\pm$0.008 & 0.017$\pm$0.008 & 0.018$\pm$0.009 \\
\hline
(3) \vbump\ $\propto$ $a\times$[Fe/H] & 1.124$\pm$0.015 & 1.221$\pm$0.021 & 1.036$\pm$0.006 \\
\dfeh\tablenotemark{2} & 0.142$\pm$0.042 & 0.131$\pm$0.039 & 0.154$\pm$0.045 \\
 & & & \\
(4) (\vbump\ $- V_{\rm TO}$) $\propto$ $a\times$[Fe/H] & 0.814$\pm$0.019 & 0.972$\pm$0.017 & 0.835$\pm$0.036 \\
\dfeh\tablenotemark{2} & 0.197$\pm$0.058 & 0.165$\pm$0.048 & 0.192$\pm$0.057 \\
\enddata
\tablenotetext{1}{The difference in helium abundance between the \cnwr\ and \cnsr.}
\tablenotetext{2}{The difference in \feh\ between the \cnwr\ and \cnwp.}
\end{deluxetable*}

\subsection{Red Giant Branch Bumps}\label{ss:bump}
During the course of low-mass star evolution, RGB stars experience a temporary drop in luminosity when the very thin H-burning shell crosses the discontinuity in the chemical composition and lowered mean molecular weight left by the deepest penetration of the convective envelope during the ascent of the RGB.
RGB stars in such evolutionary phases have to cross three times the same luminosity interval, leaving a distinctive feature, the so-called RGBB \citep[e.g., see][]{renzini88, cassisi13}.
It is well recognized that, at a given age, the RGBB luminosity increases with helium abundance and decreases with metallicity \citep[e.g., see][]{cassisi97}.
Accurate differential photometry can be possible in the GC MP study, and therefore, one can precisely estimate relative helium contents among MPs in a given GC with prior metallicity information \citep[e.g., see][]{bragaglia10, lee15, lee17, lee18, milone18b, lagioia18, lee21a}.

In order to derive the RGBB $V$ magnitudes, we calculated the generalized differential luminosity functions (LFs) using a Gaussian kernel density estimation for individual populations in 47~Tuc. We determined the peak values as our RGBB $V$ magnitudes for individual populations.
In Figure~\ref{fig:bump} and Table~\ref{tab:rgbbv}, we show our results, obtaining the RGBB $V$ magnitudes of 14.399 ($\pm$ 0.040), 14.559  ($\pm$ 0.025), 14.385  ($\pm$ 0.035), 14.492 ($\pm$0.020) mag for the \cnwp, \cnwr, \cnsp, and \cnsr, respectively. We caution that the number of RGBB stars, in particular, for the \cnwp\ population is very small and it is worrisome that our RGBB $V$ magnitude may not be accurate.

We applied the same procedures for the $UBVRI$ photometry by \citet{pbs19}, to see if their data show the same trend in the RGBB $V$ magnitudes. First, we select the proper-motion member stars using the Gaia EDR3. Then we separated the \cnw\ and \cns\ populations using the \cubi\ index,
\begin{equation}
\Delta C_{UBI} = \frac{C_{UBI} - C_{UBI,r}}{C_{UBI,r} - C_{UBI,b}},
\end{equation}
where $C_{UBI} = (U-B)-(B-I)$, while $C_{UBI,b}$ and $C_{UBI,r}$ denote the fiducials for the $C_{UBI}$-blue and $C_{UBI}$-red sequences, respectively, at a given magnitude \citep{milone12,lee19a}. To perform a populational tagging, we applied the EM algorithms for a two-component Gaussian mixture model. Through this process, we obtained populational number ratio of \nrgb\ = 32:68 ($\pm$1), consistent with our result based on the proper-motion member stars.
As already shown in Figure~\ref{fig:stetson}, the $(B-V)$ color depends more sensitively on metallicities than the helium abundances or the variations in the CNO abundances \citep[also see,][]{sbordone11}. Therefore, the $(B-V)$ color can be used as a metallicity indicator.
We calculated the $(B-V)$ color difference between individual RGB stars and the mean $(B-V)$ fiducial sequence, $\Delta(B-V)$, which shows asymmetric distributions toward the small $(B-V)$ values for both populations.
We performed the populational tagging for the \cnw\ and \cns\ populations by applying the EM algorithm for a two-component Gaussian mixture model on the $\Delta(B-V)$ distributions.
We obtained \nrgbbv\ = 6:94 ($\pm$2) for the \cnw\ population and 9:91 ($\pm$2) for the \cnw\ population, where $\Delta(B-V)_{\rm blue}$ and $\Delta(B-V)_{\rm red}$ denote RGB stars bluer or redder than the mean fiducial sequence at a given $V$ magnitude. We note that these values are consistent with those presented in Table~\ref{tab:feh} using our own photometry. Given the similar populational number ratios and the metallicity sensitivity of the $(B-V)$ color, we argue that $\Delta(B-V)_{\rm blue}$ and $\Delta(B-V)_{\rm red}$ are corresponding to the metal-poor and metal-rich subpopulations in the \cnw\ and \cns\ populations, respectively.
We calculated generalized differential LFs using a Gaussian kernel density estimation for the photometry by \citet{pbs19} and we obtained the RGBB $V$ magnitudes of 14.408 ($\pm$ 0.040), 14.530  ($\pm$ 0.025), 14.358  ($\pm$ 0.030), 14.525 ($\pm$0.020) mag for the \cnwp, \cnwr, \cnsp, and \cnsr. We show our results in Figure~\ref{fig:bump_stetson} and Table~\ref{tab:rgbbv}.
We emphasize that the RGBB $V$ magnitudes of individual populations using the photometric data by \citet{pbs19} are in good agreement with those from our results. The number of RGBB stars of the \cnwp\ population of \citet{pbs19} is still small, but it is strongly believed that the RGBB at $V$ $\sim$ 14.40 mag for the \cnwp\ population is a real feature.

In order to derive the relative helium abundances and metallicities, we compare our RGBB $V$ magnitudes with those from model isochrones.
In Figure~\ref{fig:synbump}, we show plots of the \vbump\ and (\vbumpto) against \feh\ for model isochrones with $Y$ = 0.250 (0.257), 0.275, and 0.300. We used three different isochrones currently available: Y$^2$ \citep{y2}, the PGPUC \citep{valcarce12}, and the Bag of Stellar Tracks and Isochrones \citep[BaSTI;][]{basti21}.
To determine the \vbump\ of individual model isochrones, we performed Monte Carlo simulations by constructing evolutionary population synthesis models, similar to those of our previous studies \citep[e.g., see][and references therein]{lee21a}, with various metallicity and helium contents. 
We populated 200,000 artificial stars for individual model isochrones with different chemical abundances, and we generated generalized histograms to derive the \vbump\ as we did for our observed data. We note that the \vbump\ of each model isochrone is brighter than our observations.
Therefore, we subtracted 0.17, 0.22 and 0.28 mag from the \vbump\ of the $Y^2$,  PGPUC and BaSTI model isochrones, respectively, to match observed \feh, $\sim$ $-$0.72 dex, and presumed helium abundance, $Y$ $\sim$ 0.25, for the \cnwr(RGB) population \citep[see also,][for the discrepancies in the RGBB magnitudes in model isochrones]{lee15, lee21a}.
In Table~\ref{tab:rgbbslope}, we show the results for the \vbump\ and \vbumpto\ dependencies on \feh\ and $Y$.
The difference in the RGBB $V$ magnitude of 0.067($\pm$0.032) mag between the \cnwr\ and \cnwp\ can be translated into the helium abundance difference of \dy\ = 0.029($\pm$0.014), 0.028($\pm$0.013) and 0.043($\pm$0.021) from relation (1) in Table~\ref{tab:rgbbslope} using the $Y^2$, PGPUC and BaSTI isochrones, respectively.
Note that the result from the BaSTI model isochrone is larger than the other two isochrones.

Due to the well-known discrepancy between theory and observations concerning the RGBB brightness, it was kindly recommended by the referee that comparisons of the difference in brightness between the RGBB magnitude associated with the individual populations and the corresponding theoretical quantities would be more reasonable. In this respect, we obtained the turn-off $V$ magnitude of 17.588($\pm$0.020) mag for 47~Tuc, and we calculated the (\vbumpto) of individual populations assuming they have the same turn-off magnitude. Note that this assumption may not be appropriate unless they have slightly different ages, since, at a given age, the turn-off $V$ magnitude becomes brighter with decreasing metallicity and increasing helium abundance.
Using relation (2) in Table~\ref{tab:rgbbslope}, we obtained the helium abundance difference of \dy\ = 0.016($\pm$0.008), 0.017($\pm$0.008) and 0.018($\pm$0.009) from the $Y^2$, PGPUC, and BaSTI isochrones, respectively. Note that all of three estimates provide consistent degrees of helium enhancement, but they are much smaller than those from relation (1) above.
Our estimates of helium abundance difference between the \cnwr\ and \cnsr\ are \dy\ = 0.016 -- 0.043, which are in good agreement with those of \citet[][for MS]{anderson09}, \citet[][for SGB]{dicriscienzo10}, and \citet[][for RGBB]{nataf11}, \dy\ $\sim$ 0.03. On the other hand, the relative helium abundance estimates based on the HST observations of the RGBB by \citet[][\dy\ = 0.011$\pm$0.005]{milone18b} and \citet[][\dy\ = 0.010$\pm$0.004]{lagioia18} appear to be too small compared to our results, although their estimates are marginally agree with our results from relation (2) within the uncertainties.

We also attempt to calculate the metallicity difference between the \cnwr\ and \cnwp\ based on their RGBB $V$ magnitudes. Using relation (3) in Table~\ref{tab:rgbbslope}, we obtained 0.142($\pm$0.042), 0.131($\pm$0.039) and 0.154($\pm$0.045) from the $Y^2$, PGPUC and BaSTI isochrones, respectively, all of which are slightly smaller than the metallicity difference from our \hkjwl\ photometry.
When we used relation (4), we obtained consistent results with that from the \hkjwl\ photometry, finding 0.197($\pm$0.058), 0.165($\pm$0.048) and 0.192($\pm$0.057) from the $Y^2$, PGPUC and BaSTI isochrones, respectively

As shown in Figures~\ref{fig:bump} and \ref{fig:bump_stetson}, the number of the \cnsp\ RGB stars around the RGBB regime is greater than that of the \cnwp, in particular for the case (2) (using the photometrically selected RGB stars). Therefore, it can be thought that using the \cnsp\ RGBB $V$ magnitude would be more reliable to estimate the metallicity difference between the metal-rich and metal-poor components. The RGBB $V$ magnitude of the \cnsp, $V$ = 14.385($\pm$0.035), can be interpreted in the following three ways:
\begin{itemize}
 \item[1.] The \cnsp\ population has the same helium abundance as the \cnwr\ but lower \feh.
 \item[2.] The \cnsp\ population has the same helium abundance as the \cnsr\ but lower \feh.
 \item[3.] The \cnsp\ population has the same \feh\ as the \cnwr\ and \cnsr\ but enhanced helium abundance.
\end{itemize}
In case (1) above, we obtained that the \cnsp\ population should be 0.155($\pm$0.010, relation (3)) and 0.200($\pm$0.015, relation (4)) dex lower than the \cnwr. Not surprisingly, these are the similar metallicities that we derived for the \cnwp.
In case (2), the \cnsp\ should be \dfeh\ = 0.095($\pm$0.006, relation (3)) and 0.123($\pm$0.009, relation (4)) dex more metal-poor than the \cnsr.
Finally, in case (3), the \cnsp\ should be \dy\ = 0.086($\pm$0.017, relation (1)) and 0.044($\pm$0.002, relation (2)) dex more helium enhanced than the \cnwr.

As we will show below, the RHB morphology of 47~Tuc can be explained best by case (1), i.e., the \cnsp\ RHB populations has the same helium abundance as the \cnwr\ population and a lower metallicity by \dfeh\ $\sim$ $-$0.20 dex.
Therefore, our results may suggest that both the \cnwp\ and \cnsp\ have the same metallicity and helium abundance.

\begin{figure*}
\epsscale{.85}
\figurenum{13}
\plotone{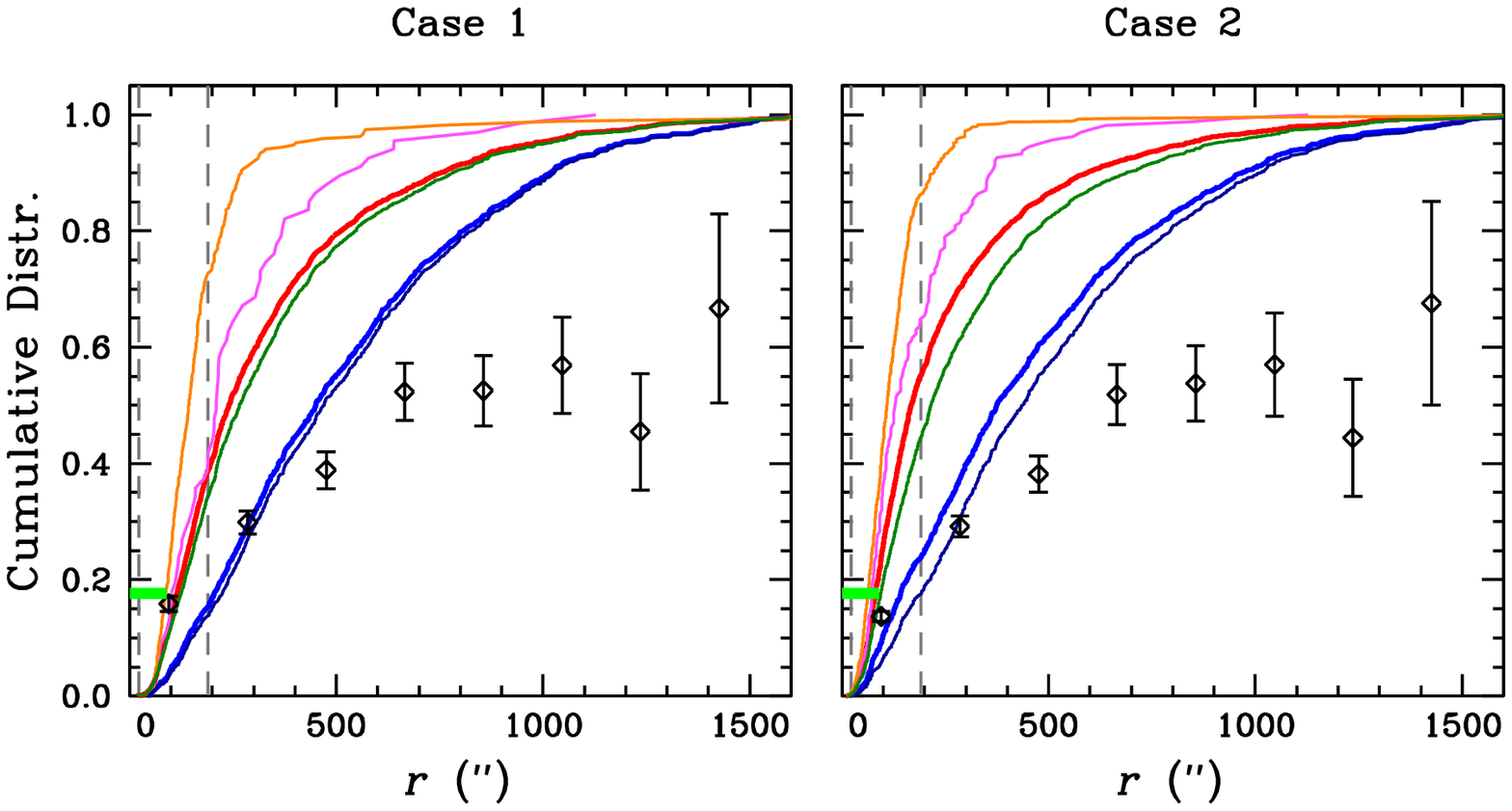}
\caption{
CRDs of individual RGB populations. Colors for individual populations are the same as Figures~\ref{fig:feh} and \ref{fig:ci}, while the thick blue and red solid lines denote the whole \cnw\ and \cns\ populations. The error bars show the fraction of the \cnw\ population with a binning radial distance of the half-light radius. The vertical gray dashed lines denote the core and the half-light radii of 47~Tuc. 
The green-shaded box indicates the fraction of the first generation by \citet{milone17}, 0.175$\pm$0.009, measured in the central part of the cluster, $\sim$ 3\arcmin\ $\times$ 3\arcmin.
}\label{fig:crd}
\end{figure*}

\begin{deluxetable*}{lcccccc}[t]
\tablenum{7}
\tablecaption{$p$ Values Returned from the K-S Tests for the CRDs of individual RGB populations\label{tab:ks}}
\tablewidth{0pc}
\tablehead{
\multicolumn{1}{c}{} &
\multicolumn{1}{c}{\cns} &
\multicolumn{1}{c}{\cnwp} &
\multicolumn{1}{c}{\cnwr} &
\multicolumn{1}{c}{\cnsp} &
\multicolumn{1}{c}{\cnsr}
}
\startdata
\cnw  & $<$  1.00$\times10^{-15}$ & $<$  1.00$\times10^{-15}$ & 0.234 & $<$  1.00$\times10^{-15}$ & $<$  1.00$\times10^{-15}$\\
\cns  & & 0.180 & $<$  1.00$\times10^{-15}$ & $<$  1.00$\times10^{-15}$ & 1.88$\times10^{-2}$ \\
\cnwp & & & $<$  1.00$\times10^{-15}$ &  2.80$\times10^{-6}$ &  3.21$\times10^{-2}$\\
\cnwr & & & &$<$  1.00$\times10^{-15}$ & $<$  1.00$\times10^{-15}$ \\
\cnsp & & & & & $<$  1.00$\times10^{-15}$\\
\enddata
\end{deluxetable*}

\subsection{Cumulative Radial Distributions}
The cumulative radial distributions (CRDs) of individual MPs may provide crucial information on the formation and dynamical evolution of MPs.
Many previous studies on the formation of GCs with MPs suggested that the \cns\ populations form in the innermost part of the cluster in a more extended \cnw\ system \citep[e.g., see][]{dercole08, bekki19, cassisi20}.
Moreover, the degree of helium enhancement or the multiphase formations of the later generation of stars can be dependent on the external gas density \citep{calura19}.
The initial structural difference between the \cnw\ and \cns\ populations can be gradually erased with time, due to the result of the preferential loss of the \cnw\ stars during the cluster's dynamical evolution \citep[e.g., see][]{vesperini21}.
It is also plausible that the current location can be affected by the stellar masses due to different degree of diffusion processes. For example, the later generations of stars with enhanced helium abundances have smaller stellar masses due to their fast evolution. Over the Hubble time, the radial distributions of the later generations of stars can expand outward \citep[e.g., see][]{fare18, calura19}.

In this context, we derive the CRDs of individual MPs in 47~Tuc, and we show our results in Figure~\ref{fig:crd}. It is very interesting to note that the metal-poor components, the \cnwp\ and \cnsp, have significantly more centrally concentrated CRDs than the metal-rich components, the \cnwr\ and \cnsr, similar to those in M3 \citep{lee21a}.
Furthermore, the \cnsp\ population is more centrally concentrated than the \cnwp\, suggesting that the metal-poor populations mimic the general trend of normal GCs, i.e., the more centrally concentrated nature of the \cns\ population.
We performed the Kolmogorov--Smirnov (K-S) tests, and our results show that the CRDs of individual populations are statistically independent each other as shown in Table~\ref{tab:ks}.

In the figure, we also show the fraction of the FG of stars (equivalent to our \cnw) from the HST photometry by \citet{milone17}, 0.175$\pm$0.009. Their FOV of HST observation was $\sim$ 3\arcmin\ $\times$ 3\arcmin, significantly smaller than that of our study, $\sim$ 1\arcdeg\ $\times$ 1\arcdeg. Due to a strong radial gradient of the populational ratio, it is most likely that \citet{milone17} underestimated the fraction of the FG of stars. We derived the fraction of the \cnw\ population within the radial distance of one half-light radius ($\sim$ 3\arcmin.2), 0.159$\pm$0.013, and our result is consistent with that of \citet{milone17}, 0.175$\pm$0.009. Our exercise shows the importance of securing a large FOV when deriving populational number ratios of GCs with MPs.

\subsection{Radial Abundance Gradients}
\begin{figure*}
\epsscale{.85}
\figurenum{14}
\plotone{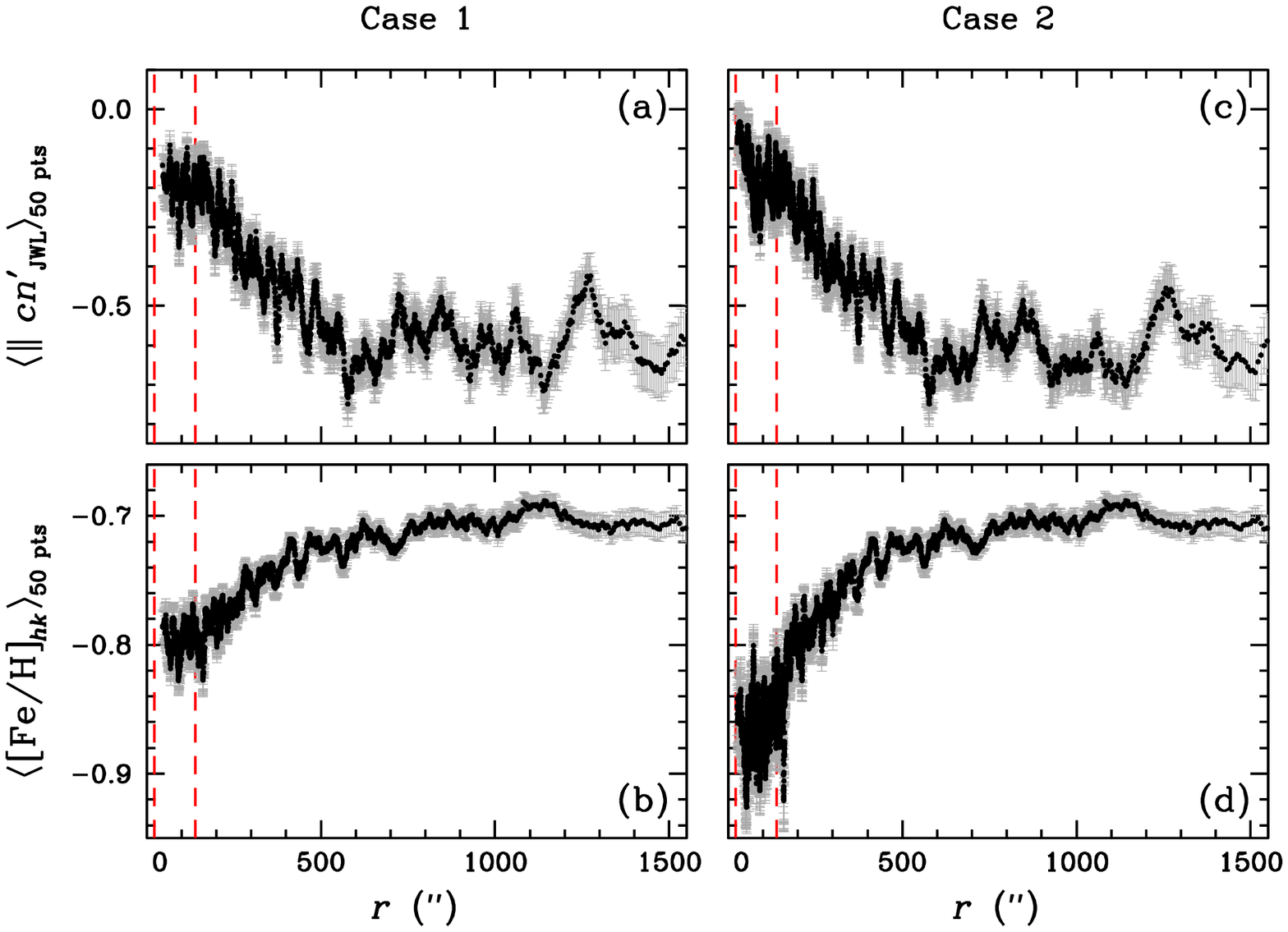}
\caption{
(a) The moving average of the adjacent 50 points for the \pcnpjwl\ index of the proper-motion member RGB stars plotted as functions of radial distance. The vertical thin gray error bars denote the standard error of the mean, and the vertical red dashed lines denote the core and the half-light radii of 47~Tuc.
(b) Same as the top panel, but for \feh.
(c) Same as (a) but for using all RGB stars.
(d) Same as (b) but for using all RGB stars.
}\label{fig:mavg}
\end{figure*}

Since the late 70s, it has been known that 47~Tuc shows the radial abundance and color gradients. For example, from their pioneering works, \citet{chun79}, and \citet{norris79} found that the \cns\ RGB stars are more centrally concentrated than the \cnw\ stars are.
Using our \cnpjwl\ index, which is a measure of the CN band strengths at \cnwave, we examined radial variation of CN strengths. We calculated the moving average of the radially adjacent 50 RGB stars, and we show our results in Figure~\ref{fig:mavg}. The figure shows that the average \cnpjwl\ value is significantly larger, i.e., more CN-rich, in the central part of the cluster, and it decreases with radial distance up to about 10\arcmin. This is consistent with the fact that the \cns\ stars are more centrally concentrated as we already discussed earlier.
When we use all RGB stars, which are photometrically selected and contain more complete samples in the central part of the cluster, the slope of the abundance gradient becomes steeper due to inclusion of more \cns\ RGB stars in the central part of the cluster.
Beyond the radial distance of 10\arcmin, the \cnpjwl\ value maintains a constant level due to the domination of the \cnw\ population in the outer regime of the cluster.

We also investigated the radial \fehhk\ variation of the cluster. As shown in Figure~\ref{fig:mavg}, the mean \fehhk\ value is smaller in the central part, and then it increases with the radial distance up to about 10\arcmin\ and maintains a constant level, consistent with the fact that the metal-poor stars are significantly more centrally concentrated than the metal-rich stars, reminiscent of M3 \citep{lee21a}.
This is also supported by the $(B-V)$ color gradient of 47~Tuc RGB stars as we showed in Figure~\ref{fig:stetson}.

\section{Red Horizontal Branch}
Since the GC RHB stars are not warm enough to completely suppress the formation of diatomic molecules, such as NH, CN, and CH, in their surface, our color indices can be useful to perform populational tagging for RHB stars \citep[][and references therein]{lee21a}.
In their pioneering work, \citet{norris82} studied RHB in 47~Tuc, finding the \cns\ RHB stars are enhanced in nitrogen by \dnfe\ $\approx$ 0.9 dex and depleted in carbon by \dcfe\ $\approx$ 0.3 dex with respect to the \cnw\ RHB stars.
They also argued the $V$ magnitude difference of 0.04 mag between the two RHB groups, and they attributed it to difference in helium abundance.

Here, we investigate the RHB population in 47~Tuc using our color indices, finding that differences not only in helium abundance but also in metallicity are necessary to explain the observed RHB CMDs.

\begin{figure}[t]
\epsscale{1.}
\figurenum{15}
\plotone{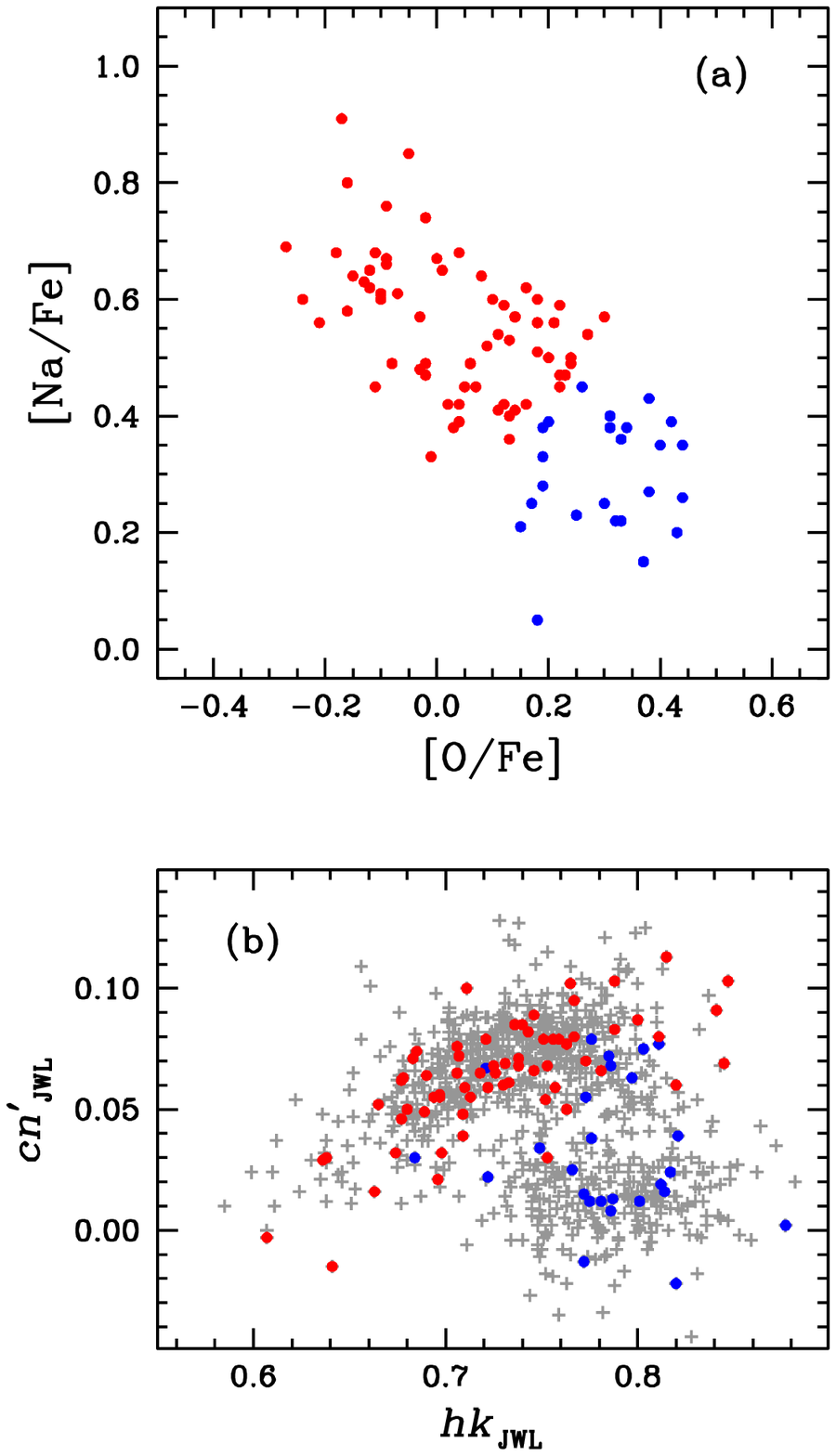}
\caption{
(a) A plot of [O/Fe] versus [Na/Fe] of the RHB stars in 47~Tuc \citep{gratton13}. The blue and red dots denote RHB stars with [Na/O] $\leq$ 0.2 (i.e., equivalent to the \cnw), and $>$ 0.2 dex (i.e., the \cns), respectively.
(b) A plot of $(b-y)$ versus \cnpjwl\ of the RHB stars in 47~Tuc along with RHB stars with [O/Fe] and [Na/Fe] measurements.
}\label{fig:hb_nao}
\end{figure}

\begin{figure}[t]
\epsscale{1.2}
\figurenum{16}
\plotone{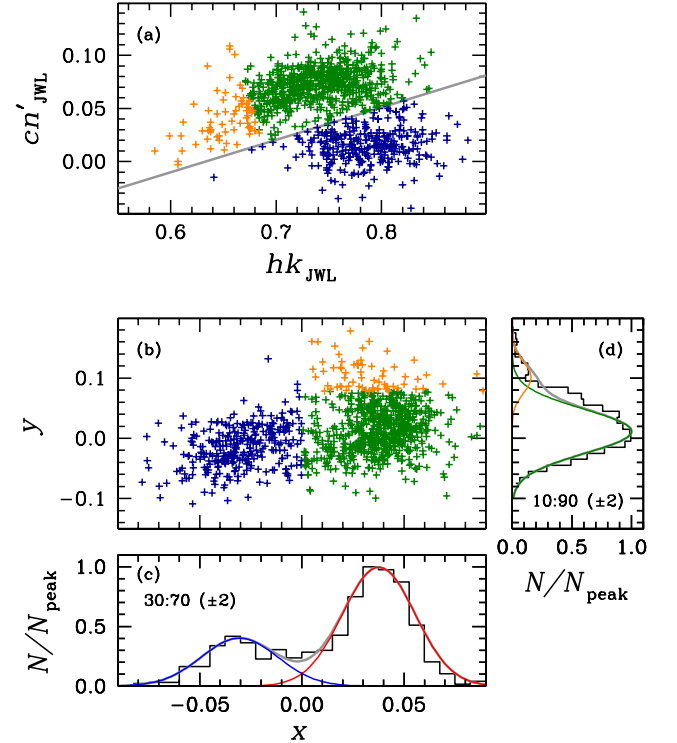}
\caption{
(a) A plot of \hkjwl\ versus \cnpjwl\ of the RHB stars in 47~Tuc, showing clustered distributions of the RHB stars. The gray solid line denotes the boundary between the \cnw(RHB) and \cns(RHB) populations. The navy, dark-green, and orange plus signs are for the \cnwhb, \cnsr(RHB), and \cnsp(RHB), respectively.
(b) The RHB distribution on the rotated plane, where the boundary between the two main bodies of the RHB stars in panel (a) is parallel to the vertical axis.
(c) The histogram of the RHB stars along the $x$-axis, showing a distinctive bimodal distribution. We also show the results returned from our EM estimator.
(d) The histogram of the \cns(RHB) along the $y$-axis, showing an asymmetric distribution. We show the results from our EM estimator
}\label{fig:hb_pop}
\end{figure}

\begin{figure*}
\epsscale{1.15}
\figurenum{17}
\plotone{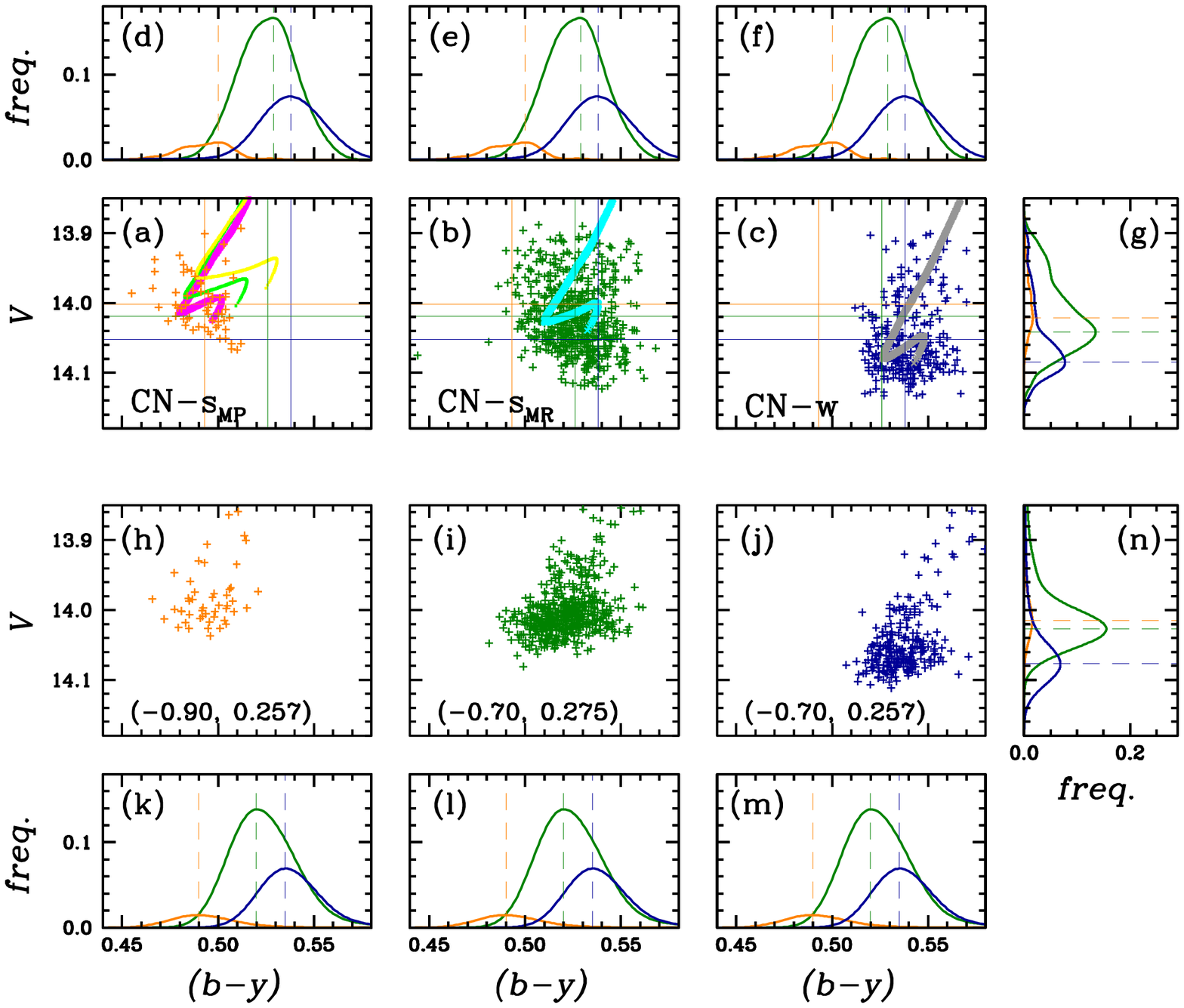}
\caption{
(a) A $(b-y)$ versus $V$ CMD of the \cnsp\ RHB stars in 47~Tuc. The vertical and horizontal thin lines are for the mean $(b-y)$ and $V$ of each population: orange (\cnsp), dark-green (\cnsr), and navy (\cnw).
The thick magenta solid line represents the BaSTI HB isochrone for (\feh, $Y$, age) = ($-$0.9, 0.257, 12.0 Gyr), while the green and yellow solid lines are for ($-$0.8, 0.275, 12.0 Gyr) and ($-$0.7, 0.300, 12.0 Gyr).
We used the distance modulus of $(m-M)_0$ = 13.21 mag and the interstellar reddening of $E(B-V)$ = 0.03 mag \citep{brogaard17}.
(b) Same as (a) but for the \cnsr\ RHB stars. The thick cyan solid line represents the BaSTI HB isochrone for ($-$0.7, 0.275, 12.0 Gyr).
(c) Same as (c) but for the \cnw\ RHB stars. The thick gray solid line represents the BaSTI HB isochrone for ($-$0.7, 0.257, 12.0 Gyr).
(d)--(f) The $(b-y)$ distributions of each population: orange (\cnsp), dark-green (\cnsr), and navy (\cnw). The dashed vertical lines denote the peak $(b-y)$ values.
(g) The $V$ distributions of each population. The dashed horizontal lines denote the peak $V$ magnitudes.
(h) Synthetic HB models for the \cnsp\ RHB stars using the BaSTI model isochrone for (\feh, $Y$, age) = ($-$0.9, 0.257, 12.0 Gyr).
(i) Same as (h) but for \cnsr\ RHB stars using the BaSTI model isochrone for ($-$0.7, 0.275, 12.0 Gyr).
(j) Same as (h) but for \cnw\ RHB stars using the BaSTI model isochrone for ($-$0.7, 0.257, 12.0 Gyr).
(k)--(m) Same as (d)--(f) but for the mean distributions of 1000 synthetic model simulation runs.
(n) Same as (g) but for the mean distributions of 1000 synthetic model simulation runs.
}\label{fig:hb_distr}
\end{figure*}

\begin{deluxetable*}{lcccc}[t]
\tablenum{8}
\tablecaption{Mean and peak RHB $V$ and $(b-y)$ from Our Observations and HB Model Simulations.\label{tab:rhbv}}
\tablewidth{0pc}
\tablehead{
\multicolumn{1}{c}{} &
\multicolumn{1}{c}{\cnw} &
\multicolumn{1}{c}{\cns} &
\multicolumn{1}{c}{\cnsp} &
\multicolumn{1}{c}{\cnsr}
}
\startdata
Mean $V$\tablenotemark{1} & 14.052$\pm$0.052$\pm$0.003 & 14.017$\pm$0.050$\pm$0.002 &
14.019$\pm$0.049$\pm$0.002 & 14.002$\pm$0.042$\pm$0.005 \\
Peak $V$\tablenotemark{1} & 14.085$\pm$0.010 & 14.039$\pm$0.010 & 14.042$\pm$0.010 & 14.021$\pm$0.010 \\
Peak $V$\tablenotemark{2} & 14.077$\pm$0.010 & 14.026$\pm$0.010 & 14.027$\pm$0.010 & 14.015$\pm$0.010 \\
Peak $V$\tablenotemark{3} & & & 13.991$\pm$0.010 & \\
Peak $V$\tablenotemark{4} & & & 13.965$\pm$0.010 & \\
\hline
Mean $(b-y)$\tablenotemark{1} & 0.538$\pm$0.015$\pm$0.001 & 0.523$\pm$0.018$\pm$0.001 &
0.526$\pm$0.015$\pm$0.001 & 0.493$\pm$0.014$\pm$0.001 \\
Peak $(b-y)$\tablenotemark{1} & 0.538$\pm$0.010 & 0.529$\pm$0.010 & 0.529$\pm$0.010 & 0.500$\pm$0.010 \\
Peak $(b-y)$\tablenotemark{2} & 0.535$\pm$0.010 & 0.520$\pm$0.010 & 0.520$\pm$0.010 & 0.490$\pm$0.010 \\
Peak $(b-y)$\tablenotemark{3} & & & 0.494$\pm$0.010 & \\
Peak $(b-y)$\tablenotemark{4} & & & 0.500$\pm$0.010 & \\
\enddata
\tablenotetext{1}{From our observations.}
\tablenotetext{2}{From synthetic HB model simulations shown in Figure~\ref{fig:hb_distr}.}
\tablenotetext{3}{From synthetic HB model simulation using the model isochrone for (\feh, $Y$, age) = ($-$0.8, 0.275, 12.0 Gyr): case (2) in \S~\ref{ss:bump}.}
\tablenotetext{4}{From synthetic HB model simulation using the model isochrone for (\feh, $Y$, age) = ($-$0.7, 0.300, 12.0 Gyr): case (3) in \S~\ref{ss:bump}.}
\end{deluxetable*}

\subsection{Populational Tagging}\label{ss:hbtag}
In Figure~\ref{fig:hb_nao}(a), we show  a plot of the \ofe\ versus \nafe\ of the RHB stars in 47~Tuc \citep{gratton13}. In the figure, the blue- and red-filled circles denote the RHB stars with [Na/O] $\leq$ 0.2 dex (the Na-normal and equivalent to the \cnw) and $>$ 0.2 dex (the Na-enhanced and equivalent to the \cns), respectively. In panel (b), we show a plot of \hkjwl\ versus \cnpjwl\ for the RHB stars in 47~Tuc. At the given \hkjwl\ values, our \cnpjwl\ index is well correlated with the sodium abundance, due to the existence of a correlation between the CN and the sodium abundances, confirming our previous result that our \cnjwl\ and \cnpjwl\ indices can be a powerful tool to investigate the MP of the RHB stars.
In the figure, the \cnpjwl\ gradient against the \hkjwl\ is partially due to the temperature effect along the RHB, in a sense that as the effective temperature of RHB stars increases the suppression of the CN band at \cnwave\ becomes greater resulting in small \cnpjwl\ values. When the effective temperature increases even higher, the influence of the H$_\zeta$ and H$_\eta$ at 3889.05 and 3835.38\AA\ becomes important, resulting in larger \cnpjwl\ values with temperatures as can be seen in the M3 RHB stars \citep[e.g., see][]{lee21a}.

We performed the populational tagging of the RHB stars in 47~Tuc using the similar method that we applied for the RGB stars. In Figure~\ref{fig:hb_pop}(a), we show a plot of \hkjwl\ versus \cnpjwl\ of the RHB stars in 47~Tuc. 
We set a boundary between the two groups of the RHB stars and rotate the \hkjwl\ versus \cnpjwl\ plane so that the boundary line becomes parallel to the vertical axis as shown in Figure~\ref{fig:hb_pop}(b). Then we applied the EM analysis with a Gaussian mixture models along the $x$-axis, which is perpendicular to the boundary line. Through this process, we obtained the populational number ratio of \nrgb\ = 30:70 ($\pm$2), consistent with that of RGB. Note that our result does not agree with those of \citet{milone12} or \citet{dondoglio21}, who employed the HST photometry and obtained \nrgb\ $\sim$ 20:80. Again, their results are based on the central part of the cluster owing to the small FOV of the HST observations. Therefore, it is natural to have a larger fraction ($\sim$80\%) of the \cns\ population due to a strong radial populational gradient of 47~Tuc.

We also examined the $y$ distributions of each RHB population. As shown in Figure~\ref{fig:hb_pop}(d), the $y$ range of the \cns(RHB) appears to be very large compared to that of the \cnw(RHB). Moreover, the $y$ distribution of the \cns(RHB) exhibits an asymmetric distribution, suggesting that the \cns(RHB) may contain multiple subpopulations as can be seen in the RGB stars.
Assuming a bimodal distribution of the \cns(RHB) population, we applied the EM analysis with a two-component Gaussian mixture model along the $y$-axis for the \cns(RHB) stars. We obtained the populational number ratio of $n$[\cnsp(RHB)]:$n$[\cnsr(RHB)] = 10:90 ($\pm$2), where the \cnsp(RHB) and \cnsr(RHB) refer the \cns\ RHB stars with high and low $y$ values, respectively.
We note that the $y$-axis is roughly the inverse of the \hkjwl\ index. Therefore, the \cnsp(RHB) stars are those with smaller \hkjwl, the photometric measure of the metallicity at a given effective temperature in the RHB temperature range.
Also it will be shown below that the \cnsp(RHB) population can be explained best with the lower metallicity by \dfeh\ $\sim$ $-$0.2 dex as can be seen in the RGB populations.

We show $(b-y)$ CMDs of the 47~Tuc RHB stars in Figure~\ref{fig:hb_distr} along with the BaSTI models \citep{basti21}, by using the distance modulus of $(m-M)_0$ = 13.21 mag and the interstellar reddening of $E(B-V)$ = 0.03 mag \citep{brogaard17}. The HB isochrones with constant mass loss for (\feh, Y, age) = ($-$0.7, 0.257, 12.0 Gyr), ($-$0.7, 0.275, 12.0 Gyr), and ($-$0.9, 0.257, 12.0 Gyr) can nicely explain the color and magnitude distributions of the \cnw(RHB), \cnsr(RHB) and \cnsp(RHB), respectively.
We note that the \cnsp(RHB) occupies about 7 ($\pm$2)\% of the total RHB stars, which is roughly consistent with that of the \cnsp\ RGB stars, $\sim$ 10 ($\pm$2)\%, within uncertainties.
On the other hand, the fraction of the \cnwp\ RGB population takes about 3\% of the total RGB stars, whose RHB counterpart we do not identify as a separate population in our analysis.
The distribution of individual RHB populations in our ($b-y$) CMD is consistent with the result of \citet{gratton13} that the sodium abundance increases as the $(B-V)$ color decreases, due to the correlation between the CN and sodium abundances in the cluster \citep[e.g.,][]{cottrell81}.

In Figure~\ref{fig:hb_distr}(a), we also show isochrones for (\feh, Y, age) = ($-$0.8, 0.275, 12.0 Gyr) and ($-$0.7, 0.300, 12.0 Gyr) to test our ideas for the cases (2) and (3) discussed in \S\ref{ss:bump}. As shown, these two RHB isochrones fail to explain the \cnsp(RHB) distribution with satisfaction, and one can rule out these two assumptions.
In Figures~\ref{fig:hb_distr}(b)--(c), we show isochrones for (\feh, Y, age) = ($-$0.7, 0.275, 12.0 Gyr) and ($-$0.7, 0.257, 12.0 Gyr), and they appear to explain  the \cnsr\ and \cnw\ RHB populations well.

We calculate the mean and peak $V$ magnitudes of individual RHB populations, and we show our results in Table~\ref{tab:rhbv}. Note that the peak magnitudes are determined from the greatest values in their smoothed distribution function with a Gaussian density estimation.
The mean color and magnitude of individual populations can be affected by the highly evolved RHB stars. It is believed that the peak color and magnitude determined by their distributions could be more practical to be compared with the model predictions as we will show below.
The \cns(RHB) stars are about 0.3 -- 0.4 mag brighter than the \cnw(RHB) stars as we already showed in Figure~\ref{fig:hb_distr}. Our results are in excellent agreement with those by \citet{norris82}, and \citet{briley97}, $\sim$ 0.04 -- 0.05 mag.

The mean $(b-y)$ color of the \cns(RHB) stars is $\sim$ 0.01 -- 0.02 mag bluer than that of the \cnw(RHB) stars, consistent with the idea that the bulk of the  \cns\ RHB stars are more helium-enhanced than the \cnw\ RHB stars are. Since the helium-enhanced RHB population will have smaller total mass and smaller ratio of the envelope mass to the total mass, they will be bluer than those with normal helium abundances \citep[e.g., see][]{cassisi20}. We also note that the range of the $(b-y)$ distribution of the \cns(RHB) stars is larger than that of the \cnw(RHB) stars, indicating that the \cns(RHB) stars experience a larger mass-loss dispersion during their RGB evolution, which seems to be odd. It could be also possible that the \cns(RHB) population contains multiple subpopulations, as we discussed for the RGB stars, the \cnsp\ and \cnsr.

To explore our idea of the existence of the metal-poor RHB component, we constructed the synthetic evolutionary population models using the (\feh, Y, age) = ($-$0.7, 0.257, 12.0 Gyr), ($-$0.7, 0.275, 12.0 Gyr), and ($-$0.9, 0.257, 12.0 Gyr) model isochrones for the \cnw, \cnsr\ and \cnsp\ populations. We performed 1000 trials to calculate the mean $(b-y)$ color and $V$ distributions for individual populations along with the peak $(b-y)$ colors and $V$ magnitudes. We show our results in Table~\ref{tab:rhbv}. Although crude, our synthetic models can reproduce the observed peak $(b-y)$ colors and $V$ magnitudes. In the table, we also show the results from the ($-$0.7, 0.300, 12.0 Gyr) and ($-$0.8, 0.275, 12.0 Gyr) models for alternative explanations of the \cnsp\ population. Our simulations with such models predict brighter peak $V$ magnitudes and slightly bluer $(b-y)$ colors than our observations. In Figure~\ref{fig:hb_distr}(h--j), we show the synthetic $(b-y)$ versus $V$ CMDs of individual population using one specific trial. We note that the distributions of synthetic models appear to be good for the \cnw\ and \cnsp\ populations. However, the wedge shape of the observed \cnsr\ population in the faint magnitude regime cannot be reproduced with a simple stellar population that we used in our study. In the future, more sophisticated HB model simulation would be very desirable \citep[e.g., see][]{salaris16}.

\begin{figure*}
\epsscale{0.85}
\figurenum{18}
\plotone{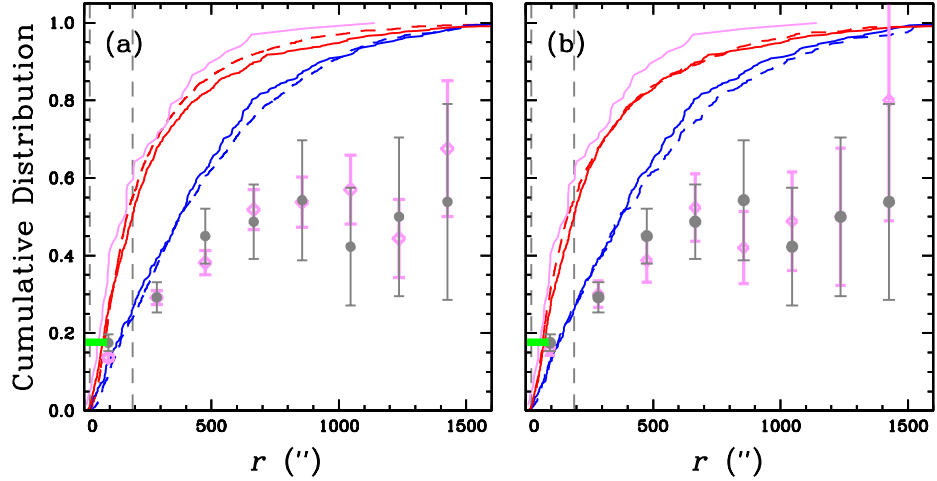}
\caption{
(a) CRDs of individual RHB populations (solid lines) along with the whole RGB populations (dashed lines). The blue, red and pink solid lines are for the \cnw(RHB), \cnsr(RHB) and \cnsp(RHB) while the blue and red dashed lines are for the \cnw(RGB) and \cns(RGB). The gray error bars show the fractions of the \cnw\ RHB population with a binning radial distance of the half-light radius, while the pink error bars show those of the \cnw\ RGB population. The vertical gray dashed lines denote the core and half-light radii of 47~Tuc. The green-shaded box indicates the fraction of the first generation by \citet{milone17} measured in the central part of the cluster.
(b) Same as (a) but for the bright RGB stars with $V \leq$ 15.0 mag.
}\label{fig:crd_hb}
\end{figure*}

\subsection{Cumulative Radial Distributions}
In Figure~\ref{fig:crd_hb}, we show the CRDs of individual RHB populations. As can be seen in the RGB, the \cns\ RHB population is more centrally concentrated than the \cnw\ RHB, which is not a surprise since the \cnw\ and \cns\ RHB stars are progenies of the \cnw\ and \cns\ RGB stars, respectively. We note that the more centrally concentrated nature of the \cns\ RHB stars was also found by \citet{milone12}.

To quantitatively examine similarity of individual CRDs, we performed the K-S tests, and we show our results is Table~\ref{tab:kshb}. When the RGB stars with the whole magnitude range of our interest, $-$1.5 $\leq$ \vvhb\ $\leq$ 2.5 mag, were used, the $p$-values related to the \cnw(RHB) population are significantly small, and it can be interpreted that none of the CRDs of the RGB populations were drawn from the same distribution of the \cnw(RHB) population. On the other hand, the \cns, \cnsr\ and \cnsp\ RHB populations show large $p$-values with the \cnwp.

It is thought that the biased sampling in our work may be responsible for this discrepancy.  RGB stars were selected from 2.5 mag fainter than the RHB stars are. At the central part of the cluster, where the \cns\ population dominates, the degree of incomplete detection of the faint stars would be great and must have affected our individual CRDs. When we restricted the bright RGB stars with $V \geq$ 15.0 mag, the agreement in individual CRDs between the RHB and RGB populations becomes improved, suggesting that each RHB population is the progeny of the corresponding RGB population.

\begin{deluxetable*}{lcccccc}[t]
\tablenum{9}
\tablecaption{$p$ Values Returned from the K-S Tests for the CRDs of individual RHB and RGB populations\label{tab:kshb}}
\tablewidth{0pc}
\tablehead{
\multicolumn{1}{c}{} &
\multicolumn{1}{c}{\cnw} &
\multicolumn{1}{c}{\cnwp} &
\multicolumn{1}{c}{\cnwr} &
\multicolumn{1}{c}{\cns}  &
\multicolumn{1}{c}{\cnsp} &
\multicolumn{1}{c}{\cnsr}
}
\startdata
\multicolumn{7}{c}{All}\\
\cnw(RHB)   & 0.026 & 3.07$\times10^{-9}$ & 0.001 & $<$ 1.00$\times10^{-15}$ & $<$ 1.00$\times10^{-15}$ & 9.03$\times10^{-9}$ \\
\cns(RHB)   & $<$ 1.00$\times10^{-15}$ & 0.425 & $<$ 1.00$\times10^{-15}$ & 2.34$\times10^{-4}$ & $<$ 1.00$\times10^{-15}$ & $<$ 1.46$\times10^{-9}$\\
\cnsp(RHB)   & 1.94$\times10^{-11}$ & 0.457 & 3.95$\times10^{-13}$ & 0.007 & 4.76$\times10^{-4}$ & 5.91$\times10^{-4}$\\
\cnsr(RHB)   & $<$ 1.00$\times10^{-15}$ & 0.383 & $<$ 1.00$\times10^{-15}$ & 0.003 & $<$ 1.00$\times10^{-15}$ & 1.04$\times10^{-7}$\\
\hline
\multicolumn{7}{c}{Bright RGB ($V \leq$ 15.0 mag)}\\
\cnw(RHB)   & 0.455 & 1.73$\times10^{-3}$ & 0.253 & $<$ 1.00$\times10^{-15}$ & $<$ 1.00$\times10^{-15}$ &  1.77$\times10^{-12}$ \\
\cns(RHB)   & $<$ 1.00$\times10^{-15}$ & 0.142 & $<$ 1.00$\times10^{-15}$ & 0.074 & $<$ 1.00$\times10^{-15}$ & 0.219\\
\cnsp(RHB)   & 9.52$\times10^{-8}$ & 0.719 & 2.31$\times10^{-8}$ & 0.668 & 1.47$\times10^{-7}$ & 0.129 \\
\cnsr(RHB)   & $<$ 1.00$\times10^{-15}$ & 0.118 & $<$ 1.00$\times10^{-15}$ & 0.041 & $<$ 1.00$\times10^{-15}$ & 0.319 \\
\enddata
\end{deluxetable*}

\section{SUMMARY}
In this paper, we presented new large FOV ($\sim$1\arcdeg$\times$1\arcdeg) Ca-CN photometry of 47~Tuc (NGC~104). We found that the \cns\ population is the major component of the cluster and it is more centrally concentrated than the \cnw\ population, similar to other Galactic GCs. We obtained the populational number ratios of the RGB and RHB: \nrgb\ = 30:70 ($\pm$1-2). Our results are in good agreement with those of previous results by others \citep[e.g.][]{nataf11, milone12}.

Following the similar method that we developed in our previous studies \citep{lee21b, lee21a}, we derived the photometric metallicity of individual RGB stars from our \hkjwl\ index, finding that both the \cnw\ and \cns\ populations show asymmetric metallicity distributions toward metal-poor regime, and they are well described by bimodal metallicity distributions. In both the \cnw\ and \cns\ populations, the metal-poor components are \dfeh\ $\sim$ 0.20 dex more metal-poor than the metal-rich components, where the metal-rich populations are main components of the clusters. The metal-poor components constitute $\sim$3\% (\cnwp) and $\sim$10\% (\cnsp) of the total RGB population. The metal-poor components (i.e., \cnwp\ and \cnsp) are significantly more centrally concentrated than the metal-rich components (i.e., \cnwr\ and \cnsr). In addition, the \cnsp\ is more centrally concentrated than the \cnwp. Therefore, the metal-poor populations themselves mimic the general trend of normal GCs that the \cns\ population is the more centrally concentrated than the \cnw.

We investigated the RGBB $V$ magnitudes, finding that $V$ = 14.399 ($\pm$ 0.040), 14.559  ($\pm$ 0.025), 14.385  ($\pm$ 0.035), 14.492 ($\pm$0.020) mag for the \cnwp, \cnwr, \cnsp, and \cnsr, respectively, which are in excellent agreement with those analyzed using the photometric data of \citet{pbs19}. We performed Monte Carlo simulations by means of the evolutionary population synthesis models, showing that the observed RGBB $V$ magnitudes are consistent with the metallicity difference of \dfeh\ $\sim$ 0.15 -- 0.20 dex between the metal-poor and metal-rich components and the helium abundance difference of \dy\ $\sim$ 0.02 -- 0.03 between the \cnwr\ and \cnsr.

Using our photometric indices, we decomposed three RHB populations in 47~Tuc. They are thought to be progenies of the \cnwr, \cnsp, and \cnsr\ based on the CN strengths and comparisons with the RHB model isochrones. Good agreements in the CRDs of RHB stars with those of the bright RGB stars may support our conclusion. It is believed that, due to its very small fraction, the RHB counterpart of the \cnwp\ cannot be clearly seen.

Recently, \citet{mckenzie21} conducted numerical simulations for the formation of a 47~Tuc-like GC in a dwarf galaxy environment. Their result indicated that newly formed GCs should be able to retain some low-mass field stars from their parent galaxy. The metal-poor components (i.e., \cnwp\ and \cnsp) that we found in this work could be field stars surrounding the \cnwr\ and \cnsr. However, the carbon and nitrogen abundance difference between the \cnwp\ and \cnsp\ cannot be easily explained in the context of the chemical evolution of the disk stars in dwarf galaxies, where the C--N anticorrelation does not appear to exist.
Instead, the metal-poor components could be understood by an independent GC system that merged with the more massive metal-rich component as we suggested for M22 and M3 \citep[e.g.,][]{lee15, lee20, lee21a}. The fraction of the metal-poor stars in our study is about 13\%, which can be translated into the total mass of $\sim$ 1.0$\times10^{4}$ \msun\  \citep{baumgardt18}.
In such a small mass system, it would be natural to expect no helium enhancement between the \cnwp\ and \cnsp\ populations \citep[e.g., see][]{milone18b}, which is exactly what we observed in 47~Tuc.

\acknowledgements
J.-W.L.\ thanks Professor Y.-C.\ Kim at Yonsei University for the discussion on the $Y^2$ isochrones and Drs. A.A.R.\ Valcarce and M.\ Catelan at Pontificia Universidad Cat\'{o}lica de Chile for reviving the PGPUC online service after the tragic fire accident that heavily affected the Institute for Astrophysics. He also thanks an anonymous referee for a careful review of the manuscript and many priceless suggestions.
J.-W.L.\ acknowledges financial support from the Basic Science Research Program (grant No.\ 2019R1A2C2086290) through the National Research Foundation of Korea (NRF) and from the faculty research fund of Sejong University in 2019.

\end{document}